\documentclass[a4paper]{article}

\usepackage{microtype}
\usepackage[american]{babel}
\usepackage{amsmath}
\usepackage{amssymb}

\usepackage[
  backend=biber,
  style=alphabetic,
  maxbibnames=99, 
  maxcitenames=2,
  useprefix=true, 
  hyperref=true,
  backref=true,
  seconds=true,
  alldates=year, 
  urldate=iso,  
  isbn=false 
]{biblatex}
\newcommand*{\Clearbibfields}{%
  \ifboolexpr{test {\ifentrytype{book}}}{}{\clearlist{publisher}}%
  \clearname{editor}%
  \ifboolexpr{test {\ifentrytype{article}}}{}{\clearfield{pages}}%
}
\AtEveryBibitem{\Clearbibfields}
\AtEveryCitekey{\Clearbibfields}

\DeclareFieldFormat{eprint:nottingham}{%
  Nottingham {\textsc{eTheses}}\addcolon\space
  \ifhyperref
    {\href{https://eprints.nottingham.ac.uk/id/eprint/#1}{\nolinkurl{#1}}}
    {\nolinkurl{#1}}}

\usepackage{authblk}
\usepackage{mathtools}
\usepackage{bussproofs}
\usepackage{cmll}
\usepackage{csquotes}
\usepackage{enumitem}
\usepackage[only=llbracket,rrbracket]{stmaryrd}
\usepackage{tikz-cd}
\usepackage{keytheorems}
\usepackage[dvipsnames]{xcolor}
\usepackage{hyperref}

\DeclareUnicodeCharacter{2202}{\ensuremath{\partial}}

\colorlet{Highlight}{purple!85!black}
\colorlet{Highlight2}{teal!95!black}
\colorlet{Url}{Aquamarine!85!black}
\colorlet{Cite}{YellowOrange!85!black}
\colorlet{Link}{violet!85!black}

\hypersetup{
  colorlinks=true,
  urlcolor=Url,
  citecolor=Cite,
  linkcolor=Link,
}

\NewDocumentCommand{\Highlight}{m}{\textcolor{Highlight}{#1}}
\NewDocumentCommand{\HighlightAlt}{m}{\textcolor{Highlight2}{#1}}

\NewDocumentCommand{\TypeTopology}{m O{}}{\href{https://www.cs.bham.ac.uk/~mhe/TypeTopology/#1.html\IfValueT{#2}{\##2}}{\texttt{#1}}}

\newkeytheoremstyle{boxed}{
  tcolorbox-no-titlebar={sharp corners=all,boxrule=1pt,colback=black!8},
}
\newkeytheoremstyle{boxed-definition}{
  inherit-style=boxed,
  numbered=false,
}
\newkeytheoremstyle{marked-definition}{
  inherit-style=definition,
  qed=$\lrcorner$,
}
\newkeytheoremstyle{construction}{
  inherit-style=definition,
  headfont={\itshape},
  qed=$\lrcorner$,
  numbered=false,
}
\newkeytheoremstyle{todo}{
  tcolorbox-no-titlebar={sharp corners=all,boxrule=1pt,colback=orange!10},
  numbered=false,
  headfont={\scshape},
}
\newkeytheorem{todo-block}[style=todo,name={Todo}]
\newkeytheorem{theorem}[style=plain,parent=section]
\newkeytheorem{
  proposition,
  lemma,
  corollary,
  problem,
  conjecture,
}[
  style=plain,
  sibling=theorem,
]
\newkeytheorem{definition}[style=marked-definition,sibling=theorem]
\newkeytheorem{construction}[style=construction,sibling=theorem]
\newkeytheorem{remark}[style=remark,sibling=theorem]
\newkeytheorem{remark*}[name=Remark,style=remark,numbered=false]
\newkeytheorem{example}[style=marked-definition,sibling=theorem]

\ExplSyntaxOn
\prop_map_inline:Nn \g__keythms_thmnames_prop
  { \cs_set_eq:cc { #1 autorefname } { #1 Autorefname } }
\ExplSyntaxOff

\AtBeginDocument{%
}

\tikzset{
  multimap/.tip={Circle[open]},
}

\title{Derivatives for Containers in Univalent Foundations}
\author{Philipp Joram}
\author{Niccol{\`o} Veltri}
\affil{Tallinn University of Technology, Estonia}
\date{}

\DeclareMathSymbol{\shortminus}{\mathbin}{AMSa}{"39}

\newcommand*{\Op}[1]{\mathsf{#1}}
\DeclareMathOperator{\Type}{\Op{Type}}
\DeclareMathOperator{\Prop}{\Op{Prop}}
\DeclareMathOperator{\Set}{\Op{Set}}
\DeclareMathOperator{\Nat}{\mathbb{N}}
\newcommand*{\WSubKern}{\mkern-2mu}
\NewDocumentCommand{\WSubOp}{m e{_}}{%
  \mathop{\kern0pt #1\IfValueT{#2}{\WSubKern}}\nolimits\IfValueT{#2}{_{#2}}%
}
\newcommand*{\W}{\WSubOp{\Op{W}}}
\DeclareMathOperator{\WRec}{\Op{W-rec}}
\DeclarePairedDelimiter{\Fin}{\lbrack}{\rbrack}
\DeclareMathOperator{\FinSet}{\Op{FinSet}}
\DeclareMathOperator{\El}{\Op{El}}
\DeclareMathOperator{\Sup}{\Op{sup}}
\DeclareMathOperator{\Dec}{\Op{Dec}}
\DeclareMathOperator{\DefEq}{\mathrel{\coloneq}}
\DeclareMathOperator{\JudgeEq}{\mathrel{\doteq}}
\DeclareMathOperator{\Fst}{\Op{fst}}

\DeclareMathOperator{\id}{\Op{id}}
\DeclareMathOperator{\Refl}{\Op{refl}}
\NewDocumentCommand{\Inv}{}{{\shortminus 1}}
\NewDocumentCommand{\Blank}{}{{\mskip0mu minus 1fill\mathunderscore\mskip1mu minus 1fill}}
\NewDocumentCommand{\Where}{}{.\,}

\DeclareMathOperator{\Fiber}{\Op{fiber}}
\DeclareMathOperator{\Singl}{\Op{singl}}
\DeclareMathOperator{\IsIsolated}{\Op{isIsolated}}

\DeclareMathOperator{\Replace}{\Op{replace}}
\DeclareMathOperator{\SigmaRemove}{\Op{\Sigma-remove}}
\DeclareMathOperator{\SigmaIsolate}{\Op{\Sigma-isolate}}

\newcommand*{\Isolated}[1]{#1^{\circ}}
\DeclareMathOperator{\Graft}{\Op{graft}}
\DeclarePairedDelimiter{\Ext}{\llbracket}{\rrbracket}
\NewDocumentCommand{\Alg}{m}{\operatorname{\Op{Alg}}_{#1}}
\DeclareMathOperator{\Rec}{\Op{rec}}

\newcommand*{\MkCont}[2]{#1\mathbin{\triangleleft}#2}
\DeclareMathOperator{\Sh}{\Op{Sh}}
\DeclareMathOperator{\Ps}{\Op{Ps}}

\NewDocumentCommand{\CTimes}{}{\mathbin{\otimes}}
\NewDocumentCommand{\CPlus}{}{\mathbin{\oplus}}
\NewDocumentCommand{\CConst}{}{\Op{k}}
\NewDocumentCommand{\CYo}{}{\Op{y}}
\DeclarePairedDelimiterXPP{\Wk}[1]%
  {} 
  {\langle} 
  {\rangle} 
  {^{\uparrow}} 
  {#1} 
\newcommand*{\Subst}[2]{#1\lbrack#2\rbrack}
\DeclareMathOperator{\List}{\Op{List}}
\DeclareMathOperator{\Bag}{\Op{Bag}}

\DeclarePairedDelimiterX{\PropTrunc}[1]%
  {\lVert} 
  {\rVert} 
  {#1} 

\newcommand*{\Cart}[2]{#1\mathrel{\multimap}#2}
\NewDocumentCommand{\CartEquiv}{}{\multimapboth}
\newcommand*{\CartIso}[2]{#1 \mathrel{\CartEquiv} #2}

\DeclareMathOperator{\Cont}{\Op{Cont}}
\NewDocumentCommand{\ContCart}{}{\operatorname{\Cont}^{{\scriptscriptstyle\multimap}}}
\NewDocumentCommand{\ICont}{m}{\Cont_{#1}}
\NewDocumentCommand{\IContCart}{m}{\ContCart_{#1}}
\NewDocumentCommand{\TruncCont}{m}{(#1)\operatorname{\Op{-Cont}}}
\NewDocumentCommand{\TruncContCart}{m}{(#1)\operatorname{\Op{-Cont}}^{{\scriptscriptstyle\multimap}}}

\NewDocumentCommand{\Der}{}{\partial}
\NewDocumentCommand{\Id}{}{\Op{Id}}
\NewDocumentCommand{\Proj}{m}{\pi_{#1}}
\DeclareMathOperator{\Chain}{\Op{chain}}
\DeclareMathOperator{\In}{\Op{in}}
\DeclareMathOperator{\Out}{\Op{out}}
\DeclareMathOperator{\MuRule}{\Op{\mu-rule}}

\DeclareMathOperator{\WIn}{\Op{W-in}}
\DeclareMathOperator{\WPathIn}{\Op{\bar{W}-in}}
\newcommand*{\WPath}{\WSubOp{\Op{\bar{W}}}}
\DeclareMathOperator{\WTop}{\Op{top}}
\DeclareMathOperator{\WBelow}{\Op{below}}

\DeclareMathOperator{\Inl}{\Op{inl}}
\DeclareMathOperator{\Inr}{\Op{inr}}

\NewDocumentCommand{\Maybe}{m}{#1 \CTimes \Id}
\DeclareMathOperator{\Nothing}{\Op{nothing}}
\DeclareMathOperator{\Just}{\Op{just}}

\begin{document}

\maketitle

\begin{abstract}
  Containers conveniently represent a wide class of inductive data types.
  Their derivatives compute representations of types of one-hole contexts,
  useful for implementing tree-traversal algorithms.
  Derivatives of discrete containers (whose positions have decidable equality) satisfy a universal property with respect to \emph{linear} morphisms of containers.
  This induces a well-behaved notion of linear substitution for one-hole contexts.

  Working in Univalent Foundations, we extend the derivative operation to \emph{untruncated} containers (whose shapes and positions are arbitrary types).
  We prove that this derivative, defined in terms of a set of \emph{isolated positions},
  satisfies an appropriate universal property in the wild category of untruncated containers and linear morphisms,
  as well as basic laws with respect to constants, sums and products.
  A chain rule exists, but is in general non-invertible.
  In fact, a globally invertible chain rule is inconsistent in the presence of non-set types,
  and equivalent to a classical principle when restricted to set-truncated containers.
  We derive a rule for derivatives of least fixpoints from the chain rule,
  and characterize its invertibility.
  All of our results are formalized in Cubical Agda.
\end{abstract}

\section{Introduction}

A container \cite{AbbottEtAl2005ContainersConstructingstrictly} encodes the signature of an inductive data type:
it consists of a collection of shapes \( S \), each \( s : S \) representing a constructor,
and for each a collection of positions \( P(s) \), indexing arguments of the constructor.
The container defines a polynomial functor \( X \mapsto \sum_{s : S} X^{P(s)} \),
called its \emph{extension}.
Natural transformations of polynomial functors are a well-studied model of parametric polymorphism~\cite{BainbridgeEtAl1990Functorialpolymorphism}.
\Citeauthor{AbbottEtAl2004ConstructingPolymorphicPrograms} show that
the extension defines an equivalence between the category of containers and that of polynomial functors and natural transformations.
This lets us reason about type-polymorphic programs entirely in the language of containers;
many operations on functors (sums, products, composition, etc.) are representable directly in terms of containers.

Another important operation on polynomial functors that has found applications in computer science is that of the \emph{derivative}.
In \citetitle{Huet1997Zipper}~\cite{Huet1997Zipper}, \citeauthor{Huet1997Zipper} describes a procedure for computing a type of contexts around a subtree in a tree-shaped data type.
\Citeauthor{McBride2001DerivativeRegularType}~\cite{McBride2001DerivativeRegularType}
notices that for the class of \emph{regular types}, one-hole contexts behave like derivatives of functions in calculus:
they satisfy similar rules for constants, sums and products, and their substitution follows a chain rule.
A regular type has a straightforward interpretation as a polynomial functor,
and \citeauthor{McBride2001DerivativeRegularType}'s computation of one-hole contexts essentially describes the derivative of such functors:
Given a polynomial \( F(X) = \sum_{s : S} X^{P(s)} \),
its derivative
\(
  \Der{F}(X) = \sum_{s : S} \sum_{p : P(s)} X^{P(s) \setminus \{ p \}}
\)
is again a polynomial functor.\footnote{%
  Notice the similarities with the analytic derivative \( \partial(\sum_i x^{b_i}) = \sum_i b_i x^{b_i - 1} \).
}

\Citeauthor{AbbottEtAl2003DerivativesContainers} make this connection precise by giving a derivative operation on containers,
first in a categorical meta-language~\cite{AbbottEtAl2003DerivativesContainers},
then in a type-theoretic one~\cite{AbbottEtAl2005DataDifferentiating}.
They show that derivatives of containers satisfy a universal property with respect to a class of \emph{linear}
morphisms of containers.
Namely, for any container \( G \) on which equality of positions is decidable,
linear morphisms \( F \to \Der{G} \) are in 1-to-1 correspondence with linear morphisms \( F \CTimes \Id \to G \).
Furthermore, the laws of derivatives are encoded as isomorphisms:
the chain rule, for example, becomes an isomorphism of containers,
\( \Der{(\Subst{F}{G})} \cong \Subst{(\Der{F})}{G} \CTimes \Der{G} \).

In the meantime, other notions of containers have appeared,
whose shapes and positions have different kinds of structure:
quotient- \cite{AbbottEtAl2004ConstructingPolymorphicPrograms} and action containers~\cite{JoramVeltri2025DataTypesSymmetries}
express symmetries by having groups act on sets of positions,
symmetric containers internalize symmetries in a groupoid of shapes~\cite{Gylterud2011},
and categorified containers~\cite{AltenkirchKaposi2021containermodeltype} describe directed relations of shapes as a category.

When working in Univalent Foundations, the most direct generalization of containers is the one from sets to arbitrary types
with potentially nontrivial higher path types:
an \emph{untruncated} container \( (\MkCont{S}{P}) \) consists of a \emph{type} of shapes \( S \), and a \emph{type} of positions \( P(s) \) for each shape \( s : S \).
Untruncated containers subsume many notions of containers that express symmetries of constructors by encoding them in the (higher) path types of shapes.
Much of the theory of set-truncated containers transfers directly, since it never assumed that types were sets in the first place.
Does the same apply to derivatives of containers?
At a glance this seems unlikely because, unlike other constructions on containers,
traditional derivatives are only well-behaved for containers whose positions have decidable equality.
By Hedberg's Theorem, positions of such \emph{discrete} containers must be sets.
As such it is not immediately clear what a derivative of containers that are genuinely not set-truncated might look like.

In this paper, we show how to define a derivative of untruncated containers in spite of this apparent obstacle.
We tackle the problem in two steps.
First, we distill a definition of derivative which depends on decidable equality in a minimal way
and does not force positions to be sets.
Then, we ask whether it behaves correctly:
Does it satisfy a universal property like discrete containers do?
Do the expected laws of derivatives hold? Is there a chain rule?

To this end, in \autoref{isolated-points} we recall the notion of \emph{isolatedness} of a point in a type,
introduced initially to prove a local version of Hedberg's Theorem:
if a point is isolated, then path spaces based at this point are \( h \)-propositions~\cite{Palmgren2011Proofrelevancefamilies}.
Consequently, for an arbitrary type \( A \), its subtype of isolated points \( \Isolated{A} \) forms an \( h \)-set.
We study how isolated points distribute over various type formers.
We make three observations:
First, isolated points distributing over \( \Sigma \)-types is a constructive taboo which would imply discreteness of arbitrary types.
Second, removal of points from a type, \( A \setminus a \), is well-behaved even for arbitrary \( A : \Type \),
as long as \( a \) is an isolated point.
Third, functions out of types pointed by an isolated point are characterized by an induction-like principle.

In \autoref{derivatives}, we use this to define a derivative that only ranges over isolated positions;
this allows us to prove the following universal property
by application of simple lemmata for isolated points:
On the (wild) category \( \ContCart \) of containers and linear morphisms,
\( \Der \) is an endofunctor,
and it is right adjoint to taking products with the identity container, \( \Maybe{\Blank} \).
As a consequence, we can solve a problem left open in \cite[14]{AbbottEtAl2005DataDifferentiating}:
the wild adjunction restricts to an ordinary adjunction on the 1-category of set-truncated containers.
In particular, this proves that \emph{every} set-truncated container has a well-behaved derivative, not just discrete ones.
Our exercise pays off: by generalizing to arbitrary types, we learn something about sets.
Verifying that this generalized derivative satisfies the basic laws for constants, sums and products is straightforward;
we essentially repurpose the proofs for the special case of discrete containers,
but distribute isolated points over type formers whenever necessary.

The goal of \autoref{chain-rule} is to derive a chain rule.
While it is possible to give a \emph{lax} chain rule, i.e. a linear morphism
from
\( \Subst{(\Der{F})}{G} \CTimes \Der{G} \)
to
\( \Der{(\Subst{F}{G})} \),
we prove that it cannot be inverted in general:
A strong chain rule for arbitrary containers is inconsistent in the presence of non-set types.
Even when restricted to set-truncated containers, this implies a constructive taboo.
We can, however, show constructively that it is an embedding of containers.

In \autoref{fixed-points}, we apply the previous results to show that derivatives of least fixpoints can be characterized entirely by the chain rule.
We ensure that every container \( F \) has a least fixpoint \( \mu F \) in the wild category of containers:
We prove that every (wild) substitution functor \( \Subst{F}{\Blank} : \ContCart \to \ContCart \) admits an initial algebra.
This is different from the usual approach in which \( \mu F \) is defined in terms of an associated container functor \( \Ext{F} \) of sets \cite{AbbottEtAl2005ContainersConstructingstrictly,DamatoEtAl2025FormalisingInductiveCoinductive}.
We do so to avoid coherence problems in the translation from container functors back to containers:
usually, properties of container functors are pulled back along \( \Ext{\Blank} \), a full and faithful functor of 1-categories.
In our setting, this is a functor of wild categories, and proving that it preserves the necessary structure is not straightforward.
Lastly, we construct a lax \( \mu \)-rule as an embedding of a fixpoint \( \mu F' \) into \( \Der(\mu F) \), in which \( F' \) is derived naturally from \( F \).
This embedding characterizes \( \Der(\mu F) \) entirely in terms of the chain rule:
it is an equivalence if and only if the chain rule for \( F \) and \( \mu F \) is an equivalence.
In other words, the \( \mu \)-rule is strong exactly when the associated chain rule is.

We develop our results as an Agda library,
implemented in its \href{https://agda.readthedocs.io/en/v2.8.0/language/cubical.html}{cubical} mode,
building on top of the \texttt{agda/cubical} library~\cite{AgdaCommunity2025CubicalAgdaLibrary}.
The library is extensively documented and contains an overview mapping all numbered results of this paper to their definitions in code.
The overview is intended to be read alongside the paper;
we therefore refrain from annotating individual statements with links into the formalization.
For convenience, we host it online at
\begin{center}
  \url{https://phijor.me/derivatives}
\end{center}
and archive an offline copy~\cite{JoramVeltri2026DerivativesContainersUnivalent}.

In this paper, we work in Homotopy Type Theory, an extension of intensional Martin--Löf type theory.
Following \AtNextCite{\defcounter{maxnames}{99}}\citeauthor{AhrensEtAl2018Categoricalstructurestype}~\cites{AhrensEtAl2018Categoricalstructurestype}{AhrensLumsdaine2019DisplayedCategories},
we may state a result as a \enquote{Problem} to highlight that it is proof-relevant,
and give a \enquote{Construction} rather than a \enquote{Proof}.
Otherwise, our notation and conventions mostly follow those of the HoTT Book~\cite{UnivalentFoundationsProgram2013HomotopyTypeTheory},
some of which we are going to recall now.

\paragraph{Types and terms.}
All of our types live in a single univalent universe, \( \Type \), unless stated otherwise.
We assume the existence of a unit type \( 1 \), an empty type \( 0 \), natural numbers \( \Nat \), products \( A \times B \), coproducts \( A + B \), and function types \( A \to B \),
as well as their dependent analogues:
For \( A : \Type \) and a family \( B : A \to \Type \),
we denote dependent function types by \( \Pi_{A} B \) or \( \prod_{a : A} B(a) \),
and dependent sum types by \( \Sigma_{A} B \) or \( \sum_{a : A} B(a) \).
We do not assume the existence of higher inductive types apart from a propositional truncation \( \PropTrunc{\Blank} \),
or where we use them to illustrate examples.
A type \( A \) is merely inhabited if \( \PropTrunc{A} \) is inhabited.
For all \( n : \Nat \), we write \( \Fin{n} : \Type \) for the standard finite type of \( n \) elements,
i.e. \( \Fin{0} \DefEq 0 \) and \( \Fin{1+n} \DefEq 1 + \Fin{n} \).

Function application is written \( f(a, b) \).
We move arguments into subscripts, \( f_a(b) \), or even drop them, \( f(b) \), if they can be inferred from context without ambiguity.
The unit type is inhabited by \( \bullet : 1 \);
the coproduct has constructors \( \Inl : A \to A + B \) and \( \Inr : B \to A + B \).
In the special case of \( A + 1 \), we call the constructors \( \Just \DefEq \Inl : A \to A + 1 \) and \( \Nothing \DefEq \Inr(\bullet) : A + 1 \).
Projections out of \( \times \) and \( \Sigma \) are denoted by \( \Op{fst} \) and \( \Op{snd} \).
We write \( \Sigma(f, g) : \Sigma_A B \to \Sigma_{A'} B' \) for the map that applies
\( f : A \to A' \) and \( g : \prod_{a : A} B(a) \to B'(f(a)) \) to the components of a \( \Sigma \)-type.

\paragraph{Equality.}
We denote by \( x \DefEq y \) a defining equality that binds a new name \( x \) to a term \( y \).
Defining equalities hold judgmentally, written \( x \JudgeEq y \), as do the usual \( \eta \)- and \( \beta \)-laws for \( \Sigma \)- and \( \Pi \)-types.
We use the same notation for \enquote{pattern matching} in proofs by induction on an inductive type.
For example, we split a proof by induction on \( x : A + B \) into cases \( x \JudgeEq \Inl(a) \) and \( x \JudgeEq \Inr(b) \).

The identity type in a type \( A \) is written \( x =_A y \), with the subscript dropped unless ambiguous.
For \( p : x = y \), we write \( p_* : B(x) \to B(y) \) for transport along \( p \) in a family \( B : A \to \Type \).
We write \( u =_p v \) for the type of dependent paths over \( p \) in \( B \) between points \( u : B(x) \) and \( v : B(y) \).

A type \( A \) is said to be contractible if \( \Op{isContr}(A) \DefEq \sum_{a : A} \prod_{b : A} a = b \) is inhabited.
It is a proposition if \( \Op{isProp}(A) \DefEq \prod_{a, b : A} a = b \) is inhabited.
We say that a type is \( n \)-truncated if it is contractible (\( n = -2 \)), a proposition (\( n = -1 \))
or its path types are \( (n-1) \)-truncated (\( n \geq 0 \)).
To avoid confusion, we name the truncation level for low \( n \):
0-truncated types are called sets, and 1-truncated types are groupoids.
If \( P : A \to \Type \) is a family of propositions, \( \Sigma_A P \) defines a subtype of \( A \).
We implicitly coerce terms \( a : \Sigma_A P \) to type \( A \) and write, for example,
\( f(a) : B \) instead of \( f(\Op{fst}(a)) \) when applying a function \( f : A \to B \).
For all \( a : A \) the \emph{singleton} type \( \Singl(a) \DefEq \sum_{b : A} a = b \) is contractible.

\paragraph{Equivalences.}
A function \( f : A \to B \) is an equivalence if for all \( y : B \) the type of fibers \( \Fiber_f(y) \DefEq \sum_{x : A} f(x) = y \) is contractible.
Being an equivalence is a proposition, and we denote the induced subtype of functions by \( A \simeq B \).

We employ various techniques to show that a given map \( f : A \to B \) is an equivalence.
In simple cases it suffices to give some \( g : B \to A \) and show that \( g \) and \( f \) are mutually inverse,
i.e.\@ that \( f \circ g = \Op{id}_B \) and \( g \circ f = \Op{id}_A \).
Whenever possible, we decompose \( f \) into intermediate steps,
after which we appeal to the 3-for-2 property: In any commutative triangle
\[
  \begin{tikzcd}[column sep=small]
      & B & \\
    A &   & C
    \ar[from=2-1,to=1-2, "g"]
    \ar[from=1-2,to=2-3, "h"]
    \ar[from=2-1,to=2-3, "f"']
  \end{tikzcd}
\]
if any two maps are equivalences, then so is the third.
From this one can derive a number of techniques that minimize the tears shed over such proofs,
best illustrated in~\cite{Jong2025FormalizingEquivalencesTears}.
Similarly, \( f \) is an equivalence if and only if it is both an embedding and a surjection:
it is a surjection if all of its fibers are merely inhabited, and an embedding (denoted \( f : A \hookrightarrow C \)) if one of the following equivalent properties holds:
\begin{itemize}
  \item \( \Fiber_f(y) \) is a proposition for all \( y : B \),
  \item \( \Fiber_f(f(x)) \) is a proposition for all \( x : A \), or
  \item \( \Op{cong}_{x,y} : x = y \to f(x) = f(y) \) is an equivalence
\end{itemize}

\paragraph{Wild categories.}
A 1-category \( \mathcal{C} \) consists of a type of objects \( \mathcal{C}_0 \) and a \emph{set} of morphisms
\( \mathcal{C}_1(x, y) \) for all \( x, y : \mathcal{C}_0 \),
together with a composition operation \( \Blank \circ \Blank \) and identity morphisms \( \Op{id} \)
satisfying the standard equalities.
A \emph{wild} category~\cite{CapriottiKraus2017UnivalentHigherCategories} consists of the same data,
except that the assumption that \( \mathcal{C}_1 \) is a family of sets is dropped.
We understand a wild category as the first approximation of an \( \infty \)-category ---
a useful tool in the absence of a fully coherent notion of higher category internal to Homotopy Type Theory:
Unlike in a 1-category, where commutativity of any diagram is a proposition,
equalities of morphisms in a wild category are \emph{data},
and we speak of them as \emph{fillers}.
When porting 1-categorical concepts to wild categories,
it is our obligation to ensure that these fillers compose coherently.

\section{Removing Points in Univalent Foundations}\label{isolated-points}

The goal of this section is to identify, for an arbitrary type, a subtype of points which can be removed from the type in a well-behaved manner.
In the classical world of sets, removal of elements is no big deal:
a set is nothing but a collection of discrete things, and discarding individual elements is done without second thought.
Types, however, model more than just discrete collections:
in Univalent Foundations, one takes the perspective that types model spaces
in which individual points are connected by potentially complicated spaces of paths.
In particular, all operations we define have to have a continuous interpretation.
Removing individual points from a space is at worst impossible to do continuously,
and at best an over-approximation of the discrete operation.

We recall the notion of isolated points of a type, and show that they are exactly the points that can be removed continuously.
Our goal is to characterize the isolated points of a number of inductive types (\autoref{isolated-points-intro}),
and to show that doing so for \( \Sigma \)-types is constructively difficult.
We demonstrate that they behave just as one would na{\"i}vely expect when removed from a type in \autoref{removing-points}.
Lastly, in \autoref{grafting}, we give an induction-like principle for types with an isolated point, characterizing functions out of it:
If \( a : A \) is isolated, then \( (A \to B) \simeq (A \setminus a \to B) \times B \)
--- any function \( f : A \to B \) is determined exactly by its restriction to \( A \setminus a \) and its value \( f(a) : B \).
This property is in general false for non-isolated points in types of higher truncation level,
since the decomposition discards the behavior of \( f \) on the higher path spaces around \( a \).

While we demonstrate properties of isolated points by freely assuming univalence and function extensionality,
we believe that a majority of them also hold with more fine-grained assumptions. 

\subsection{Isolated Points}\label{isolated-points-intro}

Recall that a type is \emph{discrete} if equality of any pair of its points is decidable.
In Martin--Löf type theory, this is a rather strong property
---
by Hedberg's Theorem, any discrete type has propositional identity types.
Under the homotopical interpretation of types, discrete types thus cannot have any interesting path structure,
and must form \( h \)-sets.

In some cases a type might be locally discrete, however.
Consider for example the sum \( A + 1 \) for an arbitrary type \( A \).
Even if equality in this type is not decidable without further assumptions on \( A \),
it should at least be possible to decide \( x = \Nothing \), no matter which \( x : A + 1 \) we are given.
After all, the constructors \( \Just \) and \( \Nothing \) are disjoint,
so case analysis should yield a decision procedure, even without looking at points of \( A \).
We call such points \emph{isolated}:%
\footnote{%
  To the best of our knowledge,
  the name \enquote{isolated} is due to \citeauthor{Voevodsky2015experimentallibraryformalized}.
  In his 2010--2011 \emph{Foundations} library~\cite{Voevodsky2015experimentallibraryformalized},
  \href{https://www.math.ias.edu/~vladimir/Foundations_library/uu0.html\#lab67}{\texttt{uu0.v}}
  has a section titled \emph{Isolated points and types with decidable equality.}
}
\begin{definition}
  A point \( a : A \) is \emph{isolated} if \( a = b \) is decidable for all \( b : A \).
  That is, isolated points satisfy the predicate \( \IsIsolated : A \to \Type \),
  \[
    \IsIsolated(a) \DefEq \prod_{b : A} \Op{Dec}(a = b)
  \]
  We denote by
  \(
    \Isolated{A} \DefEq \sum_{a : A} \IsIsolated(a)
  \)
  the subtype of isolated points.
\end{definition}

In general, a decision procedure of equality is extra structure on a type;
after all, there are possibly many ways to affirm that, \emph{yes}, \( a = b \) in an arbitrary type \( A \).
Therefore, the type \( \IsIsolated(a) \) might, \emph{a priori}, not be a proposition.
Take for example the circle \( S^1 \): there are \( \mathbb{Z} \)-many proofs of \( \Op{Dec}(\Op{base} =_{S^1} \Op{base}) \).
However, since equality has to be decidable uniformly for a point to be isolated, this does not happen:
isolated points \emph{do} have trivial path spaces.\footnote{In fact \( S^1 \) is \emph{perfect}, i.e.\@ has no isolated points at all.}
This was first noticed as a generalization of Hedberg's theorem in plain \textsc{MLTT} by \citeauthor{Palmgren2011Proofrelevancefamilies}:
\begin{lemma}[note={{\cite[Theorem~6.2]{Palmgren2011Proofrelevancefamilies}}}]\label{is-prop-isolated-path}
  If \( a : A \) is isolated, then \( a = b \) is a proposition for all \( b : A \).
\end{lemma}
An extended discussion of this result in the context of HoTT may be found in \cite[\S~3.1]{Kraus2015TruncationLevelsHomotopy}.
An important consequence of this is that being isolated is a property:
\begin{corollary}\label{is-prop-isolated-dec-path}\label{is-prop-is-isolated}
  If \( a : A \) is isolated, then \( \Dec{(a = b)} \) is a proposition for all \( b : A \).
  In particular, \( \IsIsolated(a) \) is a proposition for all \( a : A \).
\end{corollary}
Taking these together, we see that the subtype of isolated points carves out a discrete part of a type:
\begin{proposition}[note={Isolated points form a set}]\label{is-set-isolated}
  For any type \( A \), its type of isolated points \( \Isolated{A} \) is discrete, hence a set.
  \begin{proof}
    Given \( (a, h_a), (b , h_b) : \Isolated{A} \), it suffices to show that
    \( \sum_{p : a = b} h_a =_{p} h_b \) is a proposition.
    By \autoref{is-prop-isolated-path}, \( a = b \) is a proposition.
    By \autoref{is-prop-is-isolated}, both \( h_a \) and \( h_b \) are propositions,
    hence so is the dependent path type \( h_a =_{p} h_b \).
  \end{proof}
\end{proposition}

As expected, being an isolated point is stable under equivalence:
\begin{lemma}\label{is-isolated-respect-equiv}
  If \( e : A \simeq B \),
  then \( a : A \) is isolated if and only if \( e(a) : B \) is isolated.
  We write \( \Isolated{e} : \Isolated{A} \simeq \Isolated{B} \) for the induced equivalence.
\end{lemma}
In cases where a map is not an equivalence,
we might still deduce its behavior on isolated points given that it behaves \enquote{nicely} on path spaces.
This applies in particular to embeddings, which are equivalences on path spaces.
\begin{proposition}[note={Embeddings reflect isolated points}]\label{embedding-reflect-isolated}
  Let \( f : A \hookrightarrow B \) and \( a : A \).
  If \( f(a) \) is isolated in \( B \), then \( a \) is isolated in \( A \).
  \begin{proof}
    For all \( a^\prime : A \), we need to decide \( a = a^\prime \).
    By assumption, we can decide whether \( f(a) = f(a^\prime) \) or not.
    If \( f(a) \neq f(a^\prime) \), then necessarily \( a \neq a^\prime \).
    If \( f(a) = f(a^\prime) \), we get \( a = a^\prime \) since \( f \) is an embedding, which we can cancel.
  \end{proof}
\end{proposition}
In principle, we can weaken the assumptions of the previous proposition to a function \( f : A \to B \)
for which \emph{some} \( \prod_{x,y} f(x) = f(y) \to x = y \) exists -- not necessarily an inverse to \( \Op{cong}_f \):
the proof works no matter which path we pick,
and \emph{post hoc} this choice of path is unique, since it originates from an isolated point.

In the remainder, however, we will apply \autoref{embedding-reflect-isolated} exclusively to embeddings.
We can use it, for example, to show that the canonical embeddings \( \Inl : A \hookrightarrow A + B \) and \( \Inr : B \hookrightarrow A + B \)
both reflect and create isolated points:

\begin{proposition}\label{sum-embeddings-respect-isolated}
  Let \( A, B : \Type \).
  A point \( a : A \) is isolated if and only if \( \Inl{(a)} : A + B \) is isolated;
  similarly for \( b : B \) and \( \Inr{(b)} : A + B \).
  \begin{proof}
    Let \( a : A \). In the forward direction, assume that \( a \) is isolated.
    We need to show that for any \( x : A + B \), the type \( \Inl{a} = x \) is decidable.
    Consider the case of \( x \JudgeEq \Inl{a^{\prime}} \).
    We know that \( \Inl \) is an embedding, and as such there is an equivalence
    of path spaces \( (\Inl{a} = \Inl{a^\prime}) \simeq (a = a^\prime) \).
    But \( (a = a^\prime) \) is decidable by assumption,
    hence \( (\Inl{a} = \Inl{a^\prime}) \) is decidable.
    In case \( x \JudgeEq \Inr{b} \), the type \( \Inl{a} = \Inr{b} \) is empty, hence decidable.
    For the converse, apply \autoref{embedding-reflect-isolated}:
    The map \( \Inl \) is an embedding, and as such reflects isolated points.
  \end{proof}
\end{proposition}

We see that isolated points distribute over sums:
\begin{problem}\label{isolated-sum-equiv}
  Construct an equivalence \( \Isolated{(A + B)} \simeq \Isolated{A} + \Isolated{B} \).
  \begin{construction}
    Define the obvious forward and backward maps by case analysis,
    and prove that points are isolated using \autoref{sum-embeddings-respect-isolated}.
    That these maps are mutually inverse follows since being isolated is a proposition (\autoref{is-prop-is-isolated}).
  \end{construction}
\end{problem}

The point \( \bullet : 1 \) is trivially isolated, hence \( \Isolated{1} \simeq 1 \).
Consequently, \( A + 1 \) always has an isolated point, independently of the choice of \( A \):
\begin{corollary}\label{is-isolated-nothing}
  The point \( \Nothing : A + 1 \) is isolated for any type \( A \),
  and there is an equivalence \( \Isolated{(A + 1)} \simeq \Isolated{A} + 1 \).
\end{corollary}
This implements the decision procedure \( \prod_{x : A + 1} \Op{Dec}(x = \Nothing) \) that we alluded to earlier.

While it may seem obvious that the isolated points of a disjoint sum are a sum of isolated points,
describing the isolated points of \( \Sigma \)-types is a more subtle affair.
First, observe that any dependent pair of isolated points defines an isolated point in the corresponding \( \Sigma \)-type:
\begin{proposition}\label{is-isolated-pair}\label{sigma-isolate}
  Let \( A : \Type \) and \( B : A \to \Type \) with points \( a_0 : A \) and \( b_0 : B(a_0) \).
  If both \( a_0 \) and \( b_0 \) are isolated, then \( (a_0 , b_0) \) is isolated in \( \sum_{a : A} B(a) \).
  This defines a map
  \[
    \SigmaIsolate_{A,B} :
      \sum\nolimits_{a_0 : \Isolated{A}} \Isolated{B(a_0)}
        \to
      \Isolated{%
        \left(
          \sum\nolimits_{a : A} B(a)
        \right)
      }.
  \]
  \begin{proof}
    Let \( a : A \), \( b : B(a) \). Our goal is to decide whether \( (a_0 , b_0) = (a, b) \) or not.
    By extensionality of path types of dependent sums
    it suffices to decide the equivalent type \( \sum_{p : a_0 = a} b_0 = p^{\Inv}_*(b) \).
    If \( a_0 \neq a \), then this type is empty.
    Otherwise, we have some \( p : a_0 = a \), and the type is inhabited or empty
    depending on whether \( b_0 = p^{\Inv}_*(b) \) or not.
  \end{proof}
\end{proposition}

If we had an inverse to \( \SigmaIsolate \), then the converse would hold as well.
It turns out that this requirement is not only sufficient, but also necessary.
Inspecting the fibers of \( \SigmaIsolate \), we notice that they are all propositions,
hence \( \SigmaIsolate \) must be an embedding:
\begin{lemma}\label{sigma-isolate-fibers}
  For all \( A : \Type \), \( B : A \to \Type \), and \( y \JudgeEq ((a, b), \Blank) : \Isolated{(\Sigma_{A} B)} \)
  there is an equivalence \( \Op{fiber}_{\SigmaIsolate}(y) \simeq \IsIsolated(a) \times \IsIsolated(b) \).
  \begin{proof}
    By extensionality of paths in \( \Sigma \)-types, we rephrase the type of fibers over \( y \) in terms of singletons
    over \( a \) and \( b \), namely
    \[
        \textstyle
        \sum_{(a'\negthickspace, p) : \Singl(a)}
        \sum_{(b'\negthickspace, \Blank) : \Singl(p_{*}{b})}
          \IsIsolated(a') \times \IsIsolated(b')
    \]
    The claim follows by contracting away these singletons.
  \end{proof}
\end{lemma}
\begin{proposition}\label{is-embedding-sigma-isolate}
  For all \( A : \Type \) and \( B : A \to \Type \), the map \( \SigmaIsolate_{A, B} \) is an embedding.
  \begin{proof}
    By the previous result, all fibers of \( \SigmaIsolate_{A, B} \) are propositions.
  \end{proof}
\end{proposition}

A map is an equivalence if and only if it is both an embedding and a surjection,
hence we obtain the following characterization of isolated points in \( \Sigma \)-types:
\begin{lemma}\label{is-equiv-sigma-isolate-iff-isolated-pair}
  Let \( A : \Type \) and \( B : A \to \Type \).
  The following are equivalent:
  \begin{enumerate}[label=(\arabic*)]
    \item \label{is-equiv-sigma-isolate-iff-isolated-pair-is-equiv}
      \( \SigmaIsolate_{A,B} \) is an equivalence.
    \item \label{is-equiv-sigma-isolate-iff-isolated-pair-is-surj}
      \( \SigmaIsolate_{A,B} \) is a surjection.
    \item \label{is-equiv-sigma-isolate-iff-isolated-pair-pair}
      For all \( a : A \) and \( b : B(a) \),
      if \( (a , b) \) is isolated in \( \Sigma_{A} B \),
      then both \( a \) and \( b \) are isolated.
  \end{enumerate}
  \begin{proof}
    Since \( \SigmaIsolate \) is an embedding, \ref{is-equiv-sigma-isolate-iff-isolated-pair-is-equiv} and \ref{is-equiv-sigma-isolate-iff-isolated-pair-is-surj} are clearly equivalent.
    We show that \ref{is-equiv-sigma-isolate-iff-isolated-pair-is-surj} if and only if \ref{is-equiv-sigma-isolate-iff-isolated-pair-pair}:
    By \autoref{sigma-isolate-fibers}, the fiber over an arbitrary \( y \JudgeEq ((a, b), h) : \Isolated{(\Sigma_{A} B)} \) is equivalent to
    \( \IsIsolated(a) \times \IsIsolated(b) \).
    By currying, \ref{is-equiv-sigma-isolate-iff-isolated-pair-pair} holds if and only if all fibers of \( \SigmaIsolate \) are inhabited.
  \end{proof}
\end{lemma}

Sums of discrete types over a discrete base are themselves discrete,
and in this case \( \SigmaIsolate \) is trivially invertible:
\begin{corollary}\label{discrete-is-equiv-sigma-isolated}
  If \( A : \Type \) is discrete, and \( B : A \to \Type \) is a family of discrete types,
  then \( \SigmaIsolate_{A,B} \) is an equivalence.
  \begin{proof}
    Condition \ref{is-equiv-sigma-isolate-iff-isolated-pair-pair} of \autoref{is-equiv-sigma-isolate-iff-isolated-pair} is vacuously satisfied for discrete types.
  \end{proof}
\end{corollary}

Less trivially, we can state this locally for an isolated point in the base,
and obtain a generalization of \autoref{sum-embeddings-respect-isolated} from binary-
to arbitrary sums:
\begin{proposition}
  Let \( A : \Type \), \( B : A \to \Type \), and \( a : \Isolated{A} \).
  Then any \( b : B(a) \) is isolated if and only if \( (a, b) : \Sigma_{A} B \) is isolated.
  \begin{proof}
    Only the backward direction is nontrivial,
    and follows from \autoref{embedding-reflect-isolated}:
    since \( a \) is isolated,
    \( \lambda b.\, (a, b) \) is an embedding
    ---
    hence it reflects isolated points.
  \end{proof}
\end{proposition}

In general, however, it seems difficult to describe exactly when isolated points distribute this way.
In extreme cases, both \( A \) and \( B \) can be complicated types, whereas their sum \( \Sigma_A B \) is entirely trivial.
\begin{proposition}\label{discrete-iff-is-equiv-singl-isolate}
  For all types \( A \), the following are equivalent:
  \begin{enumerate}
    \item \( A \) is discrete.
    \item For all \( a_0 : A \), the map
      \[
        \SigmaIsolate_{A, a_0 = {\Blank}} : \sum_{a : \Isolated{A}} \Isolated{(a_0 = a)} \to \Isolated{\Singl(a_0)}
      \]
      is an equivalence.
  \end{enumerate}
  \begin{proof}
    If \( A \) is discrete, then so are its path types, hence \( \SigmaIsolate_{A, a_0 = \Blank} \) is an equivalence by \autoref{discrete-is-equiv-sigma-isolated}.
    In the other direction, we can show that every \( a_0 : A \) is isolated:
    Since \( \Singl(a_0) \) is a contractible type, its center \( (a_0, \Op{refl}) \) is isolated.
    Thus, by \autoref{is-equiv-sigma-isolate-iff-isolated-pair}, the first component \( a_0 \) must be isolated.
  \end{proof}
\end{proposition}
This lets us describe whether a type is discrete purely in terms of \( \SigmaIsolate \),
which will be useful in situations in which we have control over the types indexing \( \SigmaIsolate \).

\subsection{Removing Points}\label{removing-points}

For any type \( A \) and point \( a_0 : A \) we define the subtype of \enquote{\( A \) with \( a_0 \) removed}
to be \( A \setminus a_0 \DefEq \sum_{a : A} a_0 \neq a \).
Adding a point to a type \( A \) and then removing it again yields an equivalence, as expected:
\begin{problem}\label{maybe-minus-nothing-equiv}
  Define an equivalence \( (A + 1) \setminus \Nothing \simeq A \).
\end{problem}

This is an instance of the more general case where removing a point from a sum
is the same as removing it from either side, and then taking the sum:
\begin{problem}\label{sum-remove-equiv}
  Given \( A, B : \Type \), define equivalences
  \begin{align*}
    (A + B) \setminus \Inl(a_0) &\simeq (A \setminus a_0) + B \\
    (A + B) \setminus \Inr(b_0) &\simeq A + (B \setminus b_0)
  \end{align*}
  for all points \( a_0 : A \) and \( b_0 : B \), respectively.
  \begin{construction}
    Define the obvious maps in either direction by case analysis.
    These preserve inequalities since \( \Inl \) and \( \Inr \) are embeddings,
    and are inverses of each other as inequalities are always propositions.
  \end{construction}
\end{problem}

\begin{construction}[note={for \autoref{maybe-minus-nothing-equiv}}]
  The type \( (1 \setminus \bullet) \) is empty,
  so by \autoref{sum-remove-equiv}, \( (A + 1) \setminus \Nothing \simeq A + (1 \setminus \bullet) \simeq A \).
\end{construction}

Let us now consider the converse problem:
\emph{first} removing a point \( a_0 : A \), \emph{then} adding it back into the type.
Does this yield a type equivalent to \( A \)?
There is an obvious map \( \Replace_{a_0} : (A \setminus a_0) + 1 \to A \);
defined by cases
as
\(
  \Replace_{a_0}(\Just{(a, \mathunderscore)}) \DefEq a
\)
and
\(
  \Replace_{a_0}(\Nothing) \DefEq a_0
\).
Classically, this map is an isomorphism of sets;
the inverse simply maps \( a_0 \) to \( \bullet \).
In a univalent setting however, this replaces the higher path spaces of \( a_0 \)
with the trivial ones of \( \bullet \).
Consider for example the circle \( S^1 \) with \( \mathsf{base} : S^1 \).
Here \( S^1 \setminus \mathsf{base} \simeq 0 \),
hence \( (S^1 \setminus \mathsf{base}) + 1 \simeq 1 \not\simeq S^1 \):
removing \( \mathsf{base} \) removes the entire circle, which is \emph{not} contractible.

We observe that we can replicate the classical behavior exactly when \( a_0 \) is an isolated point:

\begin{proposition}\label{isolated-minus-plus-equiv}
  Let \( a_0 : A \).
  The map \( \Replace_{a_0} : (A \setminus a_0) + 1 \to A \)
  is an equivalence if and only if \( a_0 \) is isolated in \( A \).
  \begin{proof}
    First, assume that \( a_0 \) is isolated.
    We give an inverse \( g : A \to (A \setminus a_0) + 1 \) as follows.
    Define \( \Op{r} : \prod_{a : A} \Dec{(a_0 = a)} \to (A \setminus a_0) + 1 \)
    by
    \[
      \Op{r}(a, d) \DefEq
      \begin{cases}
        \Nothing & \text{if}\ d \JudgeEq \Op{yes}(\mathunderscore : a_0 = a) \\
        \Just{(a, h)} & \text{if}\ d \JudgeEq \Op{no}(h : a_0 \neq a)
      \end{cases}
    \]
    For \( i_0 : \IsIsolated(a_0) \), let \( g(a) \DefEq \Op{r}(a, i_0(a)) \).
    By \autoref{is-prop-isolated-dec-path}, \( \Dec(a_0 = a) \) (the type of \( i_0(a) \)) is a proposition for all \( a : A \);
    we use this to ensure that \( g \) computes correctly:
    \begin{align*}
      g(a_0) &\JudgeEq \Op{r}(a_0, \Highlight{i_0(a_0)}) = \Op{r}(a_0, \Highlight{\Op{yes}(\Op{refl})}) \JudgeEq \Nothing \\
      g(a)   &\JudgeEq \Op{r}(a, \Highlight{i_0(a)}) = \Op{r}(a, \Highlight{\Op{no}(h))} \JudgeEq \Just{(a, h)}
        \quad \text{where}~h : a_0 \neq a
    \end{align*}
    Hence \( {\Replace_{a_0}} \circ g = \Op{id} \) and \( g \circ {\Replace_{a_0}} = \Op{id} \).

    For the converse, assume that \( \Replace_{a_0} \) is an equivalence.
    By \autoref{is-isolated-nothing}, \( \Nothing \) is isolated in \( (A \setminus a_0 ) + 1 \),
    so \autoref{is-isolated-respect-equiv} tells us that \( \Replace_{a_0}(\Nothing) \JudgeEq a_0 \) is isolated as well.
  \end{proof}
\end{proposition}

Later on, it will become necessary to understand how to remove points from \( \Sigma \)-types.
If we think of an \( A \)-indexed sum \( \sum_{a : A} B(a) \) as a generalization of binary sums \( B_0 + B_1 \),
then we expect removal to behave similarly:
Removing some \( (a, b) \) should remove \( b : B(a) \) from the \( a \)-th
summand, leaving all other summands unchanged.
This is indeed the case, as long as we put some restrictions%
\footnote{In~\cite[\href{https://www.cs.bham.ac.uk/~mhe/TypeTopology/UF.Sets.html\#is-h-isolated}{UF.Sets}]{EscardocontributorsTypeTopology}, \( a_0 \) is called \emph{\( h \)-isolated} if \( a_0 = a_0 \) is a proposition.}
on the paths of the indexing type \( A \):
\begin{problem}\label{sigma-remove}
  Let \( A : \Type \) and \( B : A \to \Type \)
  with points \( a_0 : A \) and \( b_0 : B(a_0) \),
  and assume \( p : \Op{isProp}(a_0 = a_0) \).
  There is a map
  \[
    \operatorname{\Op{\Sigma-remove}}_p :
    \Bigl(\smashoperator{\sum_{a : A \setminus a_0}} B(a)\Bigr) + \bigl(B(a_0) \setminus b_0\bigr)
      \to
    \Bigl(\sum_{a : A} B(a) \Bigr) \setminus (a_0 , b_0)
  \]
  \begin{construction}
    We define \( \operatorname{\Op{\Sigma-remove}}_p(x) \) by cases.
    Let
    \[
      \SigmaRemove_p(\Inl{(a, h_a, b)}) \DefEq ((a, b) , h^\prime_a),
    \]
    where \( h^\prime_a : (a_0 , b_0) = (a, b) \xrightarrow{\Op{cong}_{\Op{fst}}} a_0 = a \xrightarrow{h_a} \bot \).
    In the other case, let
    \[
      \SigmaRemove_p(\Inr{(b, h_b)}) \DefEq ((a_0, b) , h^\prime_b),
    \]
    and show \( h^\prime_b : (a_0 , b_0) \neq (a_0 , b) \) as follows:
    Assume that there is some path \( p_b : (a_0 , b_0) = (a_0 , b) \).
    From this we obtain (dependent) paths \( r : a_0 = a_0 \) and \( s : b_0 =_r b \).
    Since \( a_0 = a_0 \) is a proposition, we know that \( r = \Op{refl} \),
    hence \( b_0 = b \).
    This is contradictory since we are given \( h_b : b_0 \neq b \).
  \end{construction}
\end{problem}

This map is an equivalence whenever we can decide if we are removing from a chosen index \( a_0 \):
\begin{proposition}\label{is-equiv-sigma-remove}
  Let \( A : \Type \), \( B : A \to \Type \) with \( a_0 : A \) and \( b_0 : B(a_0) \).
  If \( a_0 \) is an isolated point of \( A \), then \( \SigmaRemove \) of \autoref{sigma-remove}
  is an equivalence.
  \begin{proof}
    First note that \( a_0 = a_0 \) is a proposition by \autoref{is-prop-isolated-path},
    thus \( \SigmaRemove \) is a well-defined map.
    We construct an inverse
    \[
      \SigmaRemove^{\Inv}
        :
      \Bigl(\sum_{a : A} B(a) \Bigr) \setminus (a_0 , b_0)
        \to
      \Bigl(\smashoperator{\sum_{a : A \setminus a_0}} B(a)\Bigr) + \bigl(B(a_0) \setminus b_0\bigr)
    \]
    as follows:
    Introduce \( a : A \), \( b : B(a) \) and \( h : (a_0 , b_0) \neq (a , b) \),
    then decide whether \( a_0 = a \) or not.
    If \( p : a_0 = a \), let \( b^\prime \DefEq p^{\Inv}_*(b) : B(a_0) \) and map to \( \Inr{(b^\prime, h^\prime)} \),
    where we conclude \( h^\prime : b_0 \neq b^\prime \) from \( h \) and \( p \).
    In case that \( h : a_0 \neq a \), we map to \( \Inl{((a, h), b)} \) directly.
    It is straightforward to verify that these maps are inverses of each other.
  \end{proof}
\end{proposition}

\subsection{Grafting}\label{grafting}

Precomposing a function \( f : A \to B \) with the embedding \( \Fst : A \setminus a_0 \hookrightarrow A \) restricts it to a map \( A \setminus a_0 \to B \).
This defines an operation
\[
  \Op{detach} : \prod_{a_0 : A} (A \to B) \to (A \setminus a_0 \to B) \times B
\]
given by \( \Op{detach}_{a_0}(f) \DefEq (f \circ \Fst, f(a_0)) \).
In order to derive the chain rule for derivatives,
\citeauthor{AbbottEtAl2005DataDifferentiating}~\cite{AbbottEtAl2005DataDifferentiating} define the inverse of this operation
for \emph{discrete} \( A \):
Given a function \( g : A \setminus a_0 \to B \) and some \( b_0 : B \),
they call the process of extending \( g \) to a map \( A \to B \) \emph{grafting}.

We obtain a good notion of grafting for higher types by localizing the assumption of decidable equality:
For arbitrary types \( A, B \) and \( a_0 : A \),
\( \Op{detach}_{a_0} \) is an equivalence if \( a_0 \) is isolated in \( A \).
This defines an induction-like principle for types with a chosen isolated point \( a_0 : \Isolated{A} \):
Functions \( A \to B \) are uniquely determined by a map \( A \setminus a_0 \to B \)
and a single point \( b_0 : B \).

\begin{problem}[note={grafting}]\label{graft}\label{grafting-computation-rules}
  For types \( A \) and \( B \), construct a function
  \[
    \Graft : \prod_{a_0 : \Isolated{A}}
      \bigl((A \setminus a_0 \to B) \times B \bigr)
        \to
      (A \to B)
  \]
  that computes as follows:
  for all \( a_0 : \Isolated{A} \), \( f : A \setminus a_0 \to B \) and \( b_0 : B \),
  \begin{align*}
    \Graft_{a_0}(f, b_0)(a_0) &= b_0
      &
    &\text{and}
      &
    \adjustlimits \prod_{a : A} \prod_{h : a_0 \neq a} \Graft_{a_0}(f, b_0)(a) &= f(a, h).
  \end{align*}
\end{problem}
\begin{construction}
  Let \( a_0 : \Isolated{A} \), \( f : A \setminus a_0 \to B \) and \( b_0 : B \).
  Decide equality with \( a_0 \) to define \( \Graft_{a_0}(f, b_0) : A \to B \) as follows:
  \[
    \Graft_{a_0}(f, b_0) \DefEq
    \lambda a.\,
    \begin{cases}
      f(a, h) & \text{if}\ (h : a_0 \neq a) \\
      b_0 & \text{otherwise}
    \end{cases}
  \]
  The first computation rule is straightforward; the second follows from an application of \autoref{is-prop-isolated-dec-path}.
\end{construction}

It follows that grafting is an inverse to \( \Op{detach} \):
\begin{proposition}[note={\( \Graft \)-induction}]\label{graft-equiv}
  Let \( A, B : \Type \) and \( a_0 : A \).
  If \( a_0 \) is isolated in \( A \),
  then \( \Op{detach}_{a_0} \) is an equivalence,
  typed
  \[
    (A \setminus a_0 \to B) \times B
      \simeq
    (A \to B).
  \]
  Its inverse is given by
  \(
    \Graft_{a_0} :
    (A \setminus a_0 \to B) \times B
      \to
    (A \to B)
  \).
  \begin{proof}
    The computation rules of \autoref{grafting-computation-rules} ensure that \( \Op{detach}_{a_0} \) and \( \Graft_{a_0} \)
    are mutually inverse.
  \end{proof}
\end{proposition}

As presented here, \( \Graft \) characterizes non-dependent functions out of types with an isolated point \( a_0 : A \).
This generalizes to dependent functions, in that we can derive an equivalence
\(
  \bigl( (\prod_{a : A \setminus a_0} B(a)) \times B(a_0) \bigr) \simeq \bigl(\prod_{a : A} B(a) \bigr)
\)
for families \( B \) over \( A \).
We understand this as an \enquote{elimination operation} in the sense of \cite[{\S4}]{McBrideMcKinna2004viewleft}:
we can define arbitrary sections \( f : \prod_{a : A} B(a) \) of the \emph{motive} \( B \)
by supplying two \emph{methods} \( f^* : \prod_{a : A \setminus a_0} B(a) \) and \( b_0 : B(a_0) \).
The extra structure of \( a_0 : \Isolated{A} \)
lets us, in a sense, \enquote{pattern match} on \( a = a_0 \) and \( a \neq a_0 \).
Although not needed for later results in this paper,
we include the construction of this eliminator in our formalization of the results.

\section{Derivatives of Containers}\label{derivatives}

Containers give categorical semantics to datatypes,
in particular the inductively defined class of \emph{strictly-positive types}~\cite{Abbott2003CategoriesContainers}.
Let us recall some definitions.
\begin{definition}
  A \emph{container} \( (\MkCont{S}{P}) \) consists of \emph{shapes} \( S : \Type \) and
  a family \( P : S \to \Type \) of \emph{positions}.
  We access shapes and positions via postfix projections
  \( (\MkCont{S}{P})_{\Sh} \DefEq S \) and \( (\MkCont{S}{P})_{\Ps} \DefEq P \).
\end{definition}
A container \( (\MkCont{S}{P}) \) can be thought of as a normal form for the datatype it represents;
each shape \( s : S \) is the name of a data constructor
and each \( p : P(s) \) indexes an argument to this constructor.
If \( P(s) \) is finite, then one may think of its cardinality as the \enquote{arity} of the constructor.
We will sometimes use binders and write \( (\MkCont{s : S}{P(s)}) \) to define containers,
just like we do for \( \Sigma \) and \( \Pi \)-types.

To model (polymorphic) functions between inductive types,
one should describe how constructors of the input map to constructors in the output.
Furthermore, to each position in the output, one must point to its occurrence in the input.
\begin{definition}\label{container-morphism}
  Let \( F \JudgeEq (\MkCont{S}{P}) \) and \( G \JudgeEq (\MkCont{T}{Q}) \).
  The type of \emph{container morphisms} between \( F \) and \( G \) is
  \[
    (F \to G) \DefEq \sum_{f : S \to T} \prod_{s : S} Q_{fs} \to P_s
  \]
  We denote the shape and position components of a morphism \( {f : F \to G} \)
  by \( {f_{\Sh} : F_{\Sh} \to G_{\Sh}} \) and \( f_{\Ps} : \prod_{s : F_{\Sh}} G_{\Ps}(f_{\Sh}(s)) \to F_{\Ps}(s) \), respectively.
\end{definition}

By mapping positions from output to input it becomes possible to describe functions in which data is dropped or duplicated.
For our purposes, we have to go one step further and describe a kind of \enquote{linear} function;
one in which the mapping of positions is 1-to-1 between input and output:
\begin{definition}
  Let \( F \JudgeEq (\MkCont{S}{P}) \) and \( G \JudgeEq (\MkCont{T}{Q}) \).
  A container morphism \(f : F \to G \) is \emph{linear}
  if \( f_{\Ps}(s) : Q(f_{\Sh}(s)) \to P(s) \)
  is an equivalence of types for each \( s : S \).
  We denote the subtype of linear morphisms by \( \Cart{F}{G} \).
\end{definition}

\paragraph{Categories of containers.}
Morphisms of containers compose as expected, and together with an identity morphism \( \id_F : F \to F \)
they form a wild category \( \Cont \):
composition is associative and unital, but we make no assumption on the truncation level of the type of morphisms \( F \to G \).
A number of properties of containers only hold when this type is an \( h \)-set.
To distinguish these cases, we introduce the following subtypes of \emph{truncated} containers:
\begin{definition}
  A container \( (\MkCont{S}{P}) \) is \emph{\( (n,k) \)-truncated} if \( S \) is \( n \)-truncated,
  and \( P_s \) is \( k \)-truncated for all \( s : S \).
  A container is \emph{discrete} if \( P_s \) is a discrete type for all \( s : S \).
\end{definition}
Traditional set-based containers are \( (0,0) \)-truncated.
In this case, morphisms form an \( h \)-set.
\Citeauthor{Gylterud2011}'s \emph{symmetric containers}~\cite{Gylterud2011} are examples of containers of higher truncation level.
They have groupoids for shapes, and sets for positions, hence are exactly the \( (1,0) \)-truncated containers.

We denote by \( \ContCart \) the wide wild subcategory of \( \Cont \) containing only the linear morphisms.
Restricting to a subtype of containers,
we write \( \TruncCont{n,k} \) for the wild subcategory of \( (n,k) \)-truncated containers,
and \( \TruncContCart{n,k} \) for its restriction to linear morphisms.

\paragraph{Relation to polynomial functors.}
Each container defines a polynomial in types, that is a family \( \Type \to \Type \):
\begin{definition}\label{wild-extension}
  The \emph{extent}\footnote{also called \emph{extension} or \emph{interpretation}} of a container \( F \JudgeEq (\MkCont{S}{P}) \) is \( \Ext{F} : \Type \to \Type \)
  given by \( \Ext{F}(X) \DefEq \sum_{s : S} (P_s \to X) \).
\end{definition}
The extent of set-truncated containers defines an endofunctor of \( h \)-sets.
\Citeauthor{AbbottEtAl2005ContainersConstructingstrictly} prove that
\( \Ext{\Blank} \) fully-faithfully embeds \( \TruncCont{0,0} \) into the category of set-endofunctors.
In essence, morphisms of set-truncated containers represent exactly polymorphic functions,
that is natural transformations of polynomial endofunctors of sets~\cite[{Theorem~3.4}]{AbbottEtAl2005ContainersConstructingstrictly}.
Linear morphisms then capture the intuition of a \enquote{linear} map of datatypes,
which can only permute data, not drop or duplicate it:
The extent \( \Ext{f} : \Ext{F} \Rightarrow \Ext{G} \) of \(f : \Cart{F}{G} \) is a \emph{cartesian} natural transformation.

Furthermore, we have the following basic operations on containers:
\begin{definition}\label{container-operations}
  Given \( F \JudgeEq (\MkCont{S}{P}) \) and \( G \JudgeEq (\MkCont{T}{Q}) \), define the following:
  \begin{itemize}
    \item Product \( F \CTimes G \):
      \begin{align*}
        (F \CTimes G)_{\Sh} &\DefEq S \times T
          &
        (F \CTimes G)_{\Ps} &\DefEq \lambda (s, t).\, P_s + Q_t
      \end{align*}
    \item Sum \( F \CPlus G \):
      \begin{align*}
        (F \CPlus G)_{\Sh} &\DefEq S + T
          &
        (F \CPlus G)_{\Ps} &\DefEq
          \lambda
          \begin{cases}
            \Inl(s).\, P_s \\
            \Inr(t).\, Q_t
          \end{cases}
      \end{align*}
    \item Substitution \( \Subst{F}{G} \):
      \begin{align*}
        (\Subst{F}{G})_{\Sh} &\DefEq \sum_{s : S} (P_s \to T)
          &
        (\Subst{F}{G})_{\Ps} &\DefEq \lambda (s, f).\, \sum_{p : P_s} Q_{fp}
      \end{align*}
  \end{itemize}
  Additionally, given \( A : \Type \), we have
  \begin{itemize}
    \item a \emph{constant} \( \CConst(A) \DefEq (\MkCont{A}{0}) \),
    \item a \emph{tuple} \( \CYo(A) \DefEq (\MkCont{1}{A}) \) for all \( A : \Type \).
  \end{itemize}
  The \emph{identity container} is \( \Id \DefEq (\MkCont{1}{1}) \).
\end{definition}
For set-truncated containers,
these operations correspond exactly to (pointwise) products and sums, as well as composition, of polynomial functors.
Products and sums form a symmetric monoidal structure with units \( \CConst(1) \) and \( \CConst(0) \), respectively.
The identity container \( \Id \DefEq (\MkCont{1}{1}) \)
is a unit for substitution, which is a non-symmetric monoidal product.
The equivalence \( \Ext{\Blank} \) preserves all of these monoidal structures; it is easy to see that
\begin{align*}
  \Ext{F \CTimes G} &= \Ext{F} \times \Ext{G},
    &
  \Ext{F \CPlus G} &= \Ext{F} + \Ext{G},
    &
  \Ext{\Subst{F}{G}} &= \Ext{F} \circ \Ext{G}
\end{align*}
and that \( \Ext{\Id} \) is the identity functor.
For all \( A : \Type \), \( \Ext{\CConst{A}} \) is the constant functor \( \lambda \Blank.\, A \).
Similarly, \( \Ext{\CYo(A)}(X) \simeq (A \to X) \)
---
the composite \( \Ext{\Blank} \circ \CYo \) is essentially the Yoneda-embedding when restricted to \( h \)-sets, whence the name \( \CYo \).

\paragraph{Univalence of containers.}
The 1-category \( \TruncCont{0,0} \) of set-truncated containers is univalent.
However, neither \( \Cont \) nor \( \ContCart \) is a univalent wild category in the na{\"i}ve sense:
The canonical map taking paths of containers \( F = G \) to categorical isomorphisms\footnote{%
  that is, pairs \( f : F \to G, g : G \to F \) with chosen paths \( fg = \id_F \) and \( gf = \id_G \)
}
is in general \emph{not} an equivalence.
To compare containers in general, we thus use the following notion of \enquote{sameness} of containers:
\begin{definition}
  A morphism \( (f, u) : F \to G \) is an \emph{equivalence} of containers
  if \( f : F_{\Sh} \to G_{\Sh} \) and \( u_s : G_{\Ps}(f(s)) \to F_{\Ps}(s) \) are equivalences of types for all \( s : F_{\Sh} \).
\end{definition}
We write \( F \CartEquiv G \) for the type of equivalences of containers, as they are necessarily linear morphisms.
By two applications of univalence, \( F \CartEquiv G \) is equivalent to \( F = G \).
While we cannot prove it internally, we think of \( \Cont \) and \( \ContCart \) as \( (\infty,1) \)-categories
in which \( F \CartEquiv G \) is the \enquote{correct} notion of weak equivalence,
representing the \( \infty \)-groupoid of paths \( F = G \).
Consequently, equivalences of containers enjoy a \( \Op{J} \)-like rule, that is, the following induction principle:
\begin{problem}[note={Equivalence induction}]\label{container-J}
  Let \( G : \Cont \) and \( P : \prod_{F : \Cont} F \CartEquiv G \to \Type \).
  Define a map
  \[
    \Op{J}_{\Cont} : P(G, \id_G) \to \prod_{F : \Cont} \prod_{e : F \CartEquiv G} P(F, e)
  \]
  such that \( \Op{J}_{\Cont}(p, G, \id_G) = p \) for all \( p : P(G, \id_G) \).
  \qed
\end{problem}

This \( \Op{J} \)-rule for containers is especially useful when reasoning about types of
fillers of commutative diagrams, i.e.\@ certain types of identifications \( f_0; \ldots; f_n = g_0; \ldots; g_k \)
of container morphisms.
\begin{lemma}\label{container-precomp-equiv}
  Let \( F, G, H : \Cont \) and \( e : F \CartEquiv G \).
  The precomposition map \( e ;\Blank : (G \to H) \to (F \to H) \)
  is an equivalence of types of container morphisms.
  In particular, \( (e;h = e;h') \simeq (h = h') \) for all \( h, h' : G \to H \).
  \begin{proof}
    By application of \( \Op{J}_{\Cont} \) to \( P(F, e) \DefEq \forall H\Where \Op{isEquiv}(\lambda h : G \to H\Where e ; h) \)
    it suffices to show that \( \id_G; \Blank \) is an equivalence.
    This follows from left-unitality, \( \id_G ; h = h \).
  \end{proof}
\end{lemma}

\paragraph{Constructing container morphisms.}
When tasked to give a morphism \( f : \Cart{F}{G} \),
it is oftentimes beneficial to factor it
through (equivalent) auxiliary containers
\begin{equation*}
  \begin{tikzcd}
    F \ar[r, -multimap, "f"] & G \\
    {F\mathrlap{'}} \ar[r, -multimap, "f'"{swap}] & {G\mathrlap{'}}
    \ar[from=1-1,to=2-1, multimap-multimap]
    \ar[from=2-2,to=1-2, multimap-multimap]
  \end{tikzcd}
\end{equation*}
This lets us separate the bureaucracy of bringing \( F \) and \( G \) into a comparable shape
from the act of defining an interesting morphism \( f' : \Cart{F'}{G'} \).
In particular, \( f \) is an equivalence of containers if and only if \( f' \) is,
which is often easier to characterize.

Similarly, we implicitly use the following extensionality principle
to construct paths between container morphisms:
\begin{lemma}[note={Extensionality of container morphisms}]\label{container-morphism-ext}
  Let \( F, G : \Cont \) and \( f, g : \Cart{F}{G} \).
  The type \( f = g \) is equivalent to
  \[
    \adjustlimits\prod_{s : F_{\Sh}} \sum_{p : f_{\Sh}(s) = g_{\Sh}(s)} f_{\Ps}(s) =_p^{B} g_{\Ps}(s)
  \]
  in which the dependent path varies over the family \( B(t) \DefEq G_{\Ps}(t) \to F_{\Ps}(s) \).
  \begin{proof}
    By extensionality for paths in \( \Sigma \)- and \( \Pi \)-types, and the fact that paths of equivalences are the same as paths of their underlying functions.
  \end{proof}
\end{lemma}

\subsection{Derivatives, Universally}

The derivative of a container \( G \) represents a type of
\( G \)-shaped trees in which a chosen subtree has been removed.
For traditional containers, this can be implemented as an operation \( G \mapsto \Der{G} \) by removing a chosen position over each shape:
Given \( G \JudgeEq (\MkCont{T}{Q}) \), \citeauthor{AbbottEtAl2005DataDifferentiating} define \( \Der{G} \)
to have as shapes pairs \( (t , q) : \sum_{T} Q \), over which the positions are \( Q_t \setminus q \).
To ensure that \( \Der \) is well-behaved, it is characterized by a universal property:
on the subcategory of discrete containers, \( \Der \) extends to an endofunctor,
and induces an adjunction \( \Maybe{\Blank} \dashv \Der \).
This describes \( \Der{G} \) as \enquote{\( G \) with a hole} by a mapping-in property:
morphisms \( \Cart{F}{\Der{G}} \) are in 1-to-1 correspondence with morphisms of shape \( (f, u) : \Cart{\Maybe{F}}{G} \),
hence the position maps are equivalences \( u_s : G_{\Ps}(fs) \simeq F_{\Ps}(s) + 1 \).
In particular, there is a position \( u_s^{\Inv}(\Inr(\bullet)) : G_{\Ps}(f(s)) \) in the output
which does not correspond to a position in the input, \( F_{\Ps}(s) \), indicating the hole.

As we have seen in the previous section, removing points from a type is a subtle process in a univalent setting:
a position \( q : G_{\Ps}(t) \) is not simply a discrete point, but comes with a potentially complicated type of paths around it.
If we wanted to encode morphisms into \( \Der{G} \) by the same universal property,
we would have to avoid the entire connected component around \( q \),
that is, find some type of \enquote{hole} \( H(q) \) such that \( G_{\Ps}(f(s)) \simeq F_{\Ps}(s) + H(q) \).
But this type now depends on \( q \) and its higher path structure,
and can no longer be expressed uniformly as a simple product with the fixed container \( \Id \).

From this, we can devise two ways forward:
Either we characterize the derivative in terms of a more fine-grained universal property
that takes the dependency on higher paths into account,
or we only take derivatives with respect to positions whose path types are more uniform.
For the purpose of this paper we take the second approach,
and define a derivative in terms of \emph{isolated} positions:
\begin{definition}
  The derivative of a container \( \Der{(\MkCont{S}{P})} \DefEq (\MkCont{S'}{P'}) \)
  has shapes \( S' \DefEq \sum_{s : S} \Isolated{P_s} \) and positions \( P'(s , p) \DefEq P_s \setminus p \).
\end{definition}
\begin{example}
  Ordered lists are encoded in the container \( \List \DefEq (\MkCont{n : \Nat}{\Fin{n}}) \);
  its extent is the ordinary list functor \( \Ext{\List}(X) = \sum_{n : \Nat} X^{\Fin{n}} \).
  Intuitively, \( \Der{\List} \) computes a type equivalent to \emph{pairs of lists},
  i.e. \( \Der{\List} \CartEquiv \List \CTimes \List \).

  We compute this as follows.
  First, note that the standard finite types are discrete, hence \( \Isolated{\Fin{n}} \simeq \Fin{n} \) for all \( n : \Nat \).
  Thus, we expand
  \[
    \Der{\List} \CartEquiv \bigl( \MkCont{ (n, k) : {\textstyle \sum_{n : \Nat} \Fin{n}} }{ {\Fin{n} \setminus k} } \bigr)
  \]
  Next, observe that pairs of lists of length \( m \) and \( n \) correspond uniquely to lists of length \( m + 1 + n \) with a hole at position \( m \):
  The map \( \lambda (m, n)\Where (m + 1 + n, m) \) extends to an equivalence of type \( \Nat \times \Nat \simeq \sum_{n : \Nat} \Fin{n} \),
  which induces an equivalence
  \(
    \Fin{m + 1 + n} \setminus m \simeq \Fin{m} + \Fin{n}
  \).
  Therefore,
  \[
    \List \CTimes \List
      \CartEquiv
    \bigl( \MkCont{ (m, n) : {\Nat \times \Nat} }{ {\Fin{m} + \Fin{n}} } \bigr)
      \CartEquiv
    \bigl( \MkCont{ (n, k) : {\textstyle \sum_{n : \Nat} \Fin{n}} }{ {\Fin{n} \setminus k} } \bigr)
  \]
  and
  \( \Der{\List} \CartEquiv \List \CTimes \List \) as desired.
\end{example}

We can not only take the derivative of a container, but also act functorially on morphisms:
\begin{problem}
  Define a wild endofunctor \( \Der : \ContCart \to \ContCart \).
  That is,
  for all \( f : \Cart{F}{G} \), a morphism \( \Der{f} : \Cart{\Der{F}}{\Der{G}} \),
  such that \( \Der(\id_F) = \id_{\Der{F}} \) and \(\Der(fg) = (\Der{f})(\Der{g}) \).
  \begin{construction}
    For any \( (f, u) : \Cart{(\MkCont{S}{P})}{(\MkCont{T}{Q})} \),
    there is a canonical morphism \( \Der{(f, u)} \DefEq (f' , u') : \Cart{\Der{(\MkCont{S}{P})}}{\Der{(\MkCont{T}{Q})}} \)
    obtained as follows:
    On shapes, the map \( f' : \sum_{s : S} \Isolated{P_s} \to \sum_{t : T} \Isolated{Q_t} \)
    applies \( f \) to the first component and \( \Isolated{(u_s^\Inv)} : \Isolated{P_s} \simeq \Isolated{Q_{f(s)}} \) to the second.
    On positions, \( u'_{s,p} : Q_{f(s)} \setminus u_s^\Inv(p) \simeq P_s \setminus p \) is obtained from \( u_s \), which respects the removed point \( p \)
    (cf.\@ \autoref{is-isolated-respect-equiv}).
  \end{construction}
\end{problem}

Since isolated points always form a set, taking the derivative of a container preserves its truncation level
as long as shapes are at least sets, and positions are at least propositions:
\begin{proposition}
  For \( n \geq 0 \) and \( k \geq -1 \), the derivative of an \( (n, k) \)-truncated container is \( (n, k) \)-truncated.
  \begin{proof}
    Let \( (\MkCont{S}{P}) \) be an \( (n, k) \)-truncated container.
    By \autoref{is-set-isolated} \( \Isolated{P_s} \) is a 0-truncated type and \( S \) is \( n \)-truncated,
    thus \( \Der{(\MkCont{S}{P})}_{\Sh} \JudgeEq \sum_{s : S} \Isolated{P_s} \) is \( n \)-truncated.
    Positions are \( k \)-types since \( P_s \setminus p \) embeds into \( P_s \).
  \end{proof}
\end{proposition}

Importantly, this turns \( \Der \) into an endofunctor on the 1-category of set-truncated containers,
without having to assume that the containers are discrete:
\begin{corollary}
  Taking derivatives is a functor \( \Der : \TruncContCart{0,0} \to \TruncContCart{0,0} \).
\end{corollary}
We believe that analogues of this hold for higher truncation levels.
Symmetric containers, for example, form a bicategory \( \TruncContCart{1,0} \),
and it should be straightforward (albeit tedious) to show that \( \Der \) restricts to a pseudo\-functor on this bicategory.

We now show that this generalized derivative operation is right-adjoint to \( \Maybe{\Blank} \),
verifying that it has the desired universal property.
We define this adjunction in terms of unit and counit natural transformations,
and discuss how this relates to the original construction of \citeauthor{AbbottEtAl2005DataDifferentiating}.

\begin{problem}\label{derivative-adjunction}
  Define a wild adjunction \( (\eta, \varepsilon) : \Maybe{\Blank} \dashv \Der \),
  that is:
  \begin{enumerate}
    \item
      Two families of morphisms
      \[
        \eta : \prod_{F : \Cont} \Cart{F}{\Der{(F \CTimes \Id)}}
        \quad\text{and}\quad
        \varepsilon : \prod_{G : \Cont} \Cart{\Der{G} \CTimes \Id}{G}
      \]
    \item with fillers of naturality squares
      \begin{align*}
        &
          \begin{tikzcd}[ampersand replacement=\&]
            F
              \ar[r, -multimap, "\eta_F"]
              \ar[d, -multimap, "f"{swap}]
              \&
            \Der(\Maybe{F})
              \ar[d, -multimap, "\Der(\Maybe{f})"]
              \\
            G
              \ar[r, -multimap, "\eta_G"{swap}]
              \&
            \Der(\Maybe{G})
          \end{tikzcd}
        &
          &\text{and}
        &
        &
          \begin{tikzcd}[ampersand replacement=\&]
            \Maybe{\Der{F}}
              \ar[r, -multimap, "\varepsilon_F"]
              \ar[d, -multimap, "\Maybe{\Der{f}}"{swap}]
              \&
            F
              \ar[d, -multimap, "f"]
              \\
            \Maybe{\Der{G}}
              \ar[r, -multimap, "\varepsilon_G"{swap}]
              \&
            G
          \end{tikzcd}%
      \end{align*}
      for all \( f : \Cart{F}{G} \),
    \item and zigzag-diagrams
      \begin{align*}
        &
        \begin{tikzcd}[ampersand replacement=\&, column sep=tiny, row sep=large]
          \Maybe{F} \& \& \Maybe{F} \\
                    \& \Maybe{(\Der{(\Maybe{F})})} \& %
          \ar[from=1-1, to=1-3, -multimap, "\id"]
          \ar[from=1-1, to=2-2, -multimap, "\Maybe{{\eta_F}}"{'}]
          \ar[from=2-2, to=1-3, -multimap, "\varepsilon_{\Maybe{F}}"{'}]
        \end{tikzcd}
        &
          &\text{and}
        &
        &
        \begin{tikzcd}[ampersand replacement=\&, column sep=tiny, row sep=large]
                    \& \Der{(\Maybe{\Der{G}})} \& \\
          \Der{G} \& \& \Der{G} %
          \ar[from=2-1, to=2-3, -multimap, "\id"]
          \ar[from=2-1, to=1-2, -multimap, "\eta_{\Der{G}}"]
          \ar[from=1-2, to=2-3, -multimap, "\Der{(\varepsilon_{G})}"]
        \end{tikzcd}
      \end{align*}
  \end{enumerate}
  \begin{construction}
    Let \( F \JudgeEq (\MkCont{S}{P}) \) and define \( \eta_F : \Cart{F}{\Der{(\Maybe{F})}} \).
    On shapes, \( {\eta_F^{\Sh}} : S \to \sum_{(s, \Blank) : S \times 1} \Isolated{(P_s + 1)} \)
    sends \( s \) to \( (s, \bullet) \) and \( \mathsf{nothing} \);
    the latter is isolated by \autoref{is-isolated-nothing}.
    On positions, define \( \eta_F^{\Ps} : \prod_{s : S} (P_s + 1) \setminus \Nothing \simeq P_s \)
    as in \autoref{maybe-minus-nothing-equiv}.

    Let \( G \JudgeEq (\MkCont{T}{Q}) \); define the counit \( \varepsilon_G : \Cart{\Maybe{(\Der{G})}}{G} \) as follows:
    on shapes, \( \varepsilon_G^{\Sh} : \sum_{t : T} \Isolated{Q_t} \times 1 \to T \) is the first projection.
    On positions, the equivalence
    \(
      \varepsilon_G^{\Ps}(t , q) : Q_t \simeq (Q_t \setminus q) + 1
    \)
    is given by \autoref{isolated-minus-plus-equiv} for all \( t : T \) and \( q : \Isolated{Q_t} \).

    To construct the zigzag-fillers,
    we apply the necessary extensionality principles for functions, equivalences and sum types.
    We are left to construct paths that are almost \( \Refl \);
    only some proofs of isolation and removal need to be compared up to propositional equality.
    Construction of the naturality squares for \( \eta \) and \( \varepsilon \) is done similarly.
  \end{construction}
\end{problem}

In their original construction, \citeauthor{AbbottEtAl2005DataDifferentiating} only establish isomorphisms between the sets
\( \Cart{\Maybe{F}}{G} \) and \( \Cart{F}{\Der{G}} \), natural in \( F \).
This falls short of defining a proper adjunction since \( \Der{G} \) is left undefined for non-discrete containers \( G \).
We can, however, complete the search for a suitable subcategory of differentiable containers \cite[14]{AbbottEtAl2005DataDifferentiating}:
Our derivative is defined functorially for \emph{all} containers,
and restricting the above wild adjunction to set-truncated containers yields the following:
\begin{theorem}\label{derivative-adjunction-sets}
  In \( \TruncContCart{0,0} \), \( \Der \) is right-adjoint to tensoring \( \Blank \CTimes \Id \). \qed
\end{theorem}
From this, we can extract the familiar natural isomorphism between sets of morphisms in \( \TruncContCart{0,0} \).
In fact, the same argument lets us obtain a natural equivalence between types of morphisms for arbitrary containers,
which otherwise would be somewhat tedious to establish:
there is an equivalence \( (\Cart{F}{\Der{G}}) \simeq (\Cart{\Maybe{F}}{G}) \) natural in \( F, G : \ContCart \),
with underlying map
\begin{align*}
  {\Blank}^\sharp &: (\Cart{F}{\Der{G}}) \to (\Cart{\Maybe{F}}{G}) \\
  f^{\sharp} &\DefEq \varepsilon_G \circ (\Maybe{f})
\end{align*}
Interestingly, the proof of \cite[{Theorem 5.1}]{AbbottEtAl2005DataDifferentiating} already uses \( {\Blank}^{\sharp} \) to show that the above family of isomorphisms is natural in \( F \).

Furthermore, composing the adjunction with itself yields a notion of \( n \)-fold derivatives.
\begin{corollary}\label{nth-derivative-sets}
  For all \( n : \Nat \), there is an adjunction \( \Blank \CTimes \CYo{[n]} \dashv \Der^n \)
  in the 1-category \( \TruncContCart{0,0} \).\kern-0.2em\footnote{
    We believe that this extends to an adjunction on the entire wild category \( \ContCart \),
    but one has to carefully check that the data of the wild adjunction in \autoref{derivative-adjunction} composes coherently.
  }
\end{corollary}
Intuitively, \( \Der^n(\MkCont{S}{P}) \) encodes a type with \( n \) distinct holes:
its type of shapes is equivalent to a type of tuples \( (s, p_1, \ldots, p_n) \)
in which \( s : S \) and \( p_{i+1} : \Isolated{P_s} \setminus p_1 \setminus \ldots \setminus p_i \).
A valid position is then some \( p : \Isolated{P_s} \) that is distinct from all \( p_i \) for \( i = 1, \ldots, n \).
As such, \( \Der^n \) computes \( n \)-hole contexts.

\subsection{Basic Laws of Derivatives}

Derivatives of containers earn their name by observing how they interact with other operations on containers:
derivatives of constants are zero, derivatives of sums and products follow the familiar sum and product rules.
Let us now investigate to what extent our derivative still respects these basic laws.

It is easy to see that derivatives of constants are always zero,
and that \( \Der{\Id} \) is the constant \( \CConst(1) \).
Both factor through the following observation:
\begin{proposition}\label{derivative-prop-trunc}
  Let \( S : \Type \) and \( P : S \to \Prop \).
  There is an equivalence of containers
  \[
    { \Der{(\MkCont{S}{P})} }
      \CartEquiv
    { (\MkCont{{\textstyle \sum_{S} P }}{0}) }
  \]
  In particular, we have
  \( \Der{(\Id)} \CartEquiv \CConst(1) \)
  and \( \Der(\CConst(A)) \CartEquiv \CConst(0) \) for all \( A : \Type \).
  \begin{proof}
    Since \( P_s \) is a proposition, we know that \( \Isolated{P_s} \simeq P_s \) and \( P_s \setminus p \simeq 0 \).
    Thus,
    \begin{align*}
      \Der{(\MkCont{S}{P})}
        &\CartEquiv (\MkCont{((s, p) : \textstyle\sum_{s : S} \Isolated{P_s})}{P_s \setminus p})
          \\
        &\CartEquiv (\MkCont{((s, p) : \textstyle\sum_{s : S} P_s)}{0})
          \qedhere
    \end{align*}
  \end{proof}
\end{proposition}

Similarly, we convince ourselves that \( \Der \) distributes over (binary) sums,
and that derivatives of products follow a Leibniz rule:
\begin{proposition}\label{sum-product-rule}\label{sum-rule}\label{leibniz-rule}
  For containers \( F, G \), the following hold:
  \begin{enumerate}
    \item Sum rule: \( {\Der{(F \CPlus G)}} \CartEquiv {\Der{F} \CPlus \Der{G}} \)
    \item Leibniz rule: \( {\Der{(F \CTimes G)}} \CartEquiv {(\Der{F} \CTimes G) \CPlus (F \CTimes \Der{G})} \)
  \end{enumerate}
  \begin{proof}
    Let \( F \JudgeEq (\MkCont{S}{P}) \) and \( G \JudgeEq (\MkCont{T}{Q}) \).
    Both equivalences are established as in the discrete setting
    (cf.~\cite[{Proposition 6.3 and 6.4}]{AbbottEtAl2005DataDifferentiating}),
    with one exception:
    to derive the Leibniz rule,
    one needs to show that isolated points distribute over binary sums in
    \[
      \sum_{s : S} \sum_{t : T} \Isolated{(P_s + Q_t)}
        \simeq
      \sum_{s : S} \sum_{t : T} \Isolated{P_s} + \Isolated{Q_t},
    \]
    which is done via \autoref{isolated-sum-equiv}.
  \end{proof}
\end{proposition}

To some extent we can also solve differential equations involving \( \Der \).
In particular, given a container \( F \), we can ask if it has an anti-derivative,
i.e. some \( G \) for which \( \CartIso{\Der{G}}{F} \).
Interestingly, \( \Der \) has fixpoints, that is containers that are their own derivative.
The prototypical example of such a fixpoint is the container of finite multisets, or \emph{bags}.
Famously, bags cannot be expressed as set-truncated containers, but can be if one allows shapes of higher truncation level.
Let us reproduce \citeauthor{Gylterud2011}'s construction~\cite[{Example~3.6.1}]{Gylterud2011} of bags as a symmetric container,
but in the language of HoTT.
Recall that a type is considered finite
if there is some \( n : \Nat \) for which it is merely equivalent to \( \Fin{n} : \Type \).%
\footnote{Specifically, such types are called Bishop-finite~\cite[{Definition~4.4}]{FruminEtAl2018FiniteSetsHomotopy}.}
The universe of finite sets,
\( \FinSet \DefEq \sum_{X : \Type} \sum_{n : \Nat} \lVert X \simeq \Fin{n} \rVert \),
comes with a map \( \El \DefEq \Fst : \FinSet \to \Type \) projecting out the underlying type.
Together, these form the container of \emph{bags}, \( \Bag \DefEq (\MkCont{\FinSet}{\El}) \).
As written, the shapes of this container quantify over all types, hence live in a higher universe.
It is possible to replace \( \FinSet \) by a higher inductive type in \( \Type \)~\cite[Theorem~25]{FinsterEtAl2021CartesianBicategoryPolynomial};
the present form, however, makes the following immediate:
\begin{proposition}\label{bag-fixed-point}
  \( \Bag \) is a \( \Der \)-fixpoint:
  there is an equivalence \( \CartIso{\Der{\Bag}}{\Bag} \).
  \begin{proof}
    First, note that finite sets are closed under addition and removal of points:
    if \( X \) is finite, then so are \( X + 1 \) and \( X \setminus x \) for all \( x : X \).
    On shapes, we construct an equivalence
    \(
      \sum_{X : \FinSet} \Isolated{\El(X)} \simeq \FinSet
    \)
    from mutually inverse functions \( f \) and \( g \).
    From left to right, define \( f(X, x_0) \DefEq X \setminus x_0 \);
    the other way let \( g(X) \DefEq (X + 1 , \Nothing) \).
    By univalence, finite sets are equal if their carrier types are equivalent,
    so \( f \) and \( g \) are inverses by \autoref{isolated-minus-plus-equiv} and \autoref{maybe-minus-nothing-equiv}.
    Given \( X : \FinSet \) and \( x_0 : \El(X) \), positions are related by the identity equivalence,
    that is
    \(
      \Bag_{\Ps}(f(X, x_0)) = \El(f(X, x_0)) = X \setminus x_0 = \Der{\Bag}_{\Ps}(X, x_0)
    \).
  \end{proof}
\end{proposition}

Unlike in classical analysis, where the exponential function is the unique solution to
the differential equation \( f^{\prime} = f \) with initial condition \( f(0) = 1 \),
the situation for containers is more nuanced:
While \( \Bag \) is a solution for \( \CartIso{\Der{F}}{F} \) such that \( { F[\CConst{0}] \CartEquiv \CConst{1} } \),
it is far from being the only one.
This is not entirely unexpected:
containers are closely related to Joyal's \emph{combinatorial species}
(as discussed e.g.~in Yorgey's thesis \cite[67]{Yorgey2014CombinatorialSpeciesLabelled}),
and these are known to have many non-isomorphic solutions even for simple differential equations,
as shown by Labelle in~\cite{Labelle1986combinatorialdifferentialequations}.

Modulo size issues, the proof of \autoref{bag-fixed-point} goes through for any subuniverse
of types closed under addition and removal of single points:
\begin{proposition}
  Let \( P : \Type \to \Prop \) be a predicate
  such that for all \( A : \Type \), \( P(A) \) implies both \( P(A + 1) \)
  and \( \prod_{a : A} P(A \setminus a) \).
  This defines a container
  \( \Bag_P \DefEq (\MkCont{\sum_{\Type} P}{\Fst}) \)
  such that \( \CartIso{\Der{\Bag_P}}{\Bag_P} \). \qedhere
  \begin{proof}
    The proof mirrors that of \autoref{bag-fixed-point};
    for the details we refer the reader to the formalization.
  \end{proof}
\end{proposition}
\( \Bag \) is an instance of the above for the predicate \( P(A) \DefEq \sum_{n : \Nat} \PropTrunc{A \simeq \Fin{n}} \).
In the formalization, we provide another example:
for \( P(A) \DefEq \lVert A \hookrightarrow \Nat \rVert \),
\( \Bag_P \) is a container representing countable multisets,
and again a \( \Der{} \)-fixpoint.

\section{The Chain Rule}\label{chain-rule}

In addition to the basic properties of the previous section,
we expect the derivative to satisfy an analogue of the chain rule
\( (f \circ g)^{\prime} = (f^{\prime} \circ g) \cdot g^{\prime} \),
which describes how derivatives distribute over composition.
In our setting, substitution \( \Subst{\Blank}{\Blank} \) takes on the role of composition,
and for discrete containers, \citeauthor{AbbottEtAl2005DataDifferentiating} show that
\( \Der{\Subst{F}{G}} \) is indeed isomorphic to \( \Subst{(\Der{F})}{G} \CTimes \Der{G} \).
Na{\"\i}vely attempting to lift their proof to untruncated containers,
we unfortunately run into difficulties:
While it is possible to define a morphism from one to the other,
it is not immediately clear that an inverse exists.
Precisely, we obtain the following \emph{directed} or \emph{lax} chain rule:
\begin{problem}[note={The lax chain rule}]\label{lax-chain-rule}
  For any two containers \( F, G \), define a morphism
  \[
    \Chain_{F,G} :
    \Cart%
      { \Subst{(\Der{F})}{G} \CTimes \Der{G} }%
      {\Der{(\Subst{F}{G})}}
  \]
\end{problem}
\begin{construction}
  Let \( F \JudgeEq (\MkCont{S}{P}) \) and \( G \JudgeEq (\MkCont{T}{Q}) \).
  As usual, we have to construct a map on shapes and an equivalence of positions.
  On shapes, our goal is a map
  \begin{multline*}
    {\Chain_{F,G}^{\Sh}} :
      \sum\nolimits_{(s, p) : \sum_{s : S} \Isolated{P_s}} (P_s \setminus p \to T)
        \times
      \sum\nolimits_{t : T} \Isolated{Q_t}
      \\
    \to
      \sum\nolimits_{(s, f) : \sum_{s : S} (P_s \to T)} \Isolated{\left( \sum\nolimits_{p : P_s} Q_{fp} \right)}
  \end{multline*}
  Let us first reshape its domain by some equivalences.
  By reassociating the sums, we obtain
  \begin{align*}
    &\mathrel{\hphantom{\simeq}}
      \sum\nolimits_{(s, p) : \sum_{s : S} \Isolated{P_s}} (P_s \setminus p \to T)
        \times
      \sum_{t : T} \Isolated{Q_t}
    \\
    &\simeq
      \sum\nolimits_{(s, p) : \sum_{s : S} \Isolated{P_s}} \sum\nolimits_{(f, t) : (P_s \setminus p \to T) \times T} \Isolated{Q_t}
  \intertext{%
    The induction principle for \( \Graft \) tells us that the types \( (P_s \setminus p \to T) \times T \) and \( P_s \to T \) are equivalent (\autoref{graft-equiv}),
    thus we simplify to
  }
    &\simeq
      \sum\nolimits_{(s, p) : \sum_{s : S} \Isolated{P_s}} \sum\nolimits_{f : P_s \to T} \Isolated{Q_{fp}}
  \intertext{%
    By permuting the sum yet again, we are left with
  }
    &\simeq
      \sum\nolimits_{(s, f) : \sum_{s : S} \,(P_s \to T)} \bigl( \sum_{p : \Isolated{P_s}} \Isolated{(Q_{fp})} \bigr)
  \end{align*}
  Denote this equivalence by \( \lambda \).
  Now, the left and the right only differ in
  \begin{align*}
    \sum_{p : \Isolated{P_s}} \Isolated{(Q_{fp})}
      \qquad\text{vs.}\qquad
    \Isolated{\bigl( \sum_{p : P_s} Q_{fp} \bigr)}
  \end{align*}
  \Autoref{sigma-isolate} gives us a map
  \(
    \SigmaIsolate_{P_s,Q_{f(\Blank)}} :
      \sum_{p : \Isolated{P_s}} \Isolated{(Q_{fp})}
        \to
      \Isolated{\bigl( \sum_{p : P_s} Q_{fp} \bigr)}
  \),
  hence
  \(
    \cramped{
      \Chain_{F,G}^{\Sh} \DefEq
        \Sigma(\id, \SigmaIsolate_{P_s,Q_{f(\Blank)}}) \circ \lambda
    }
  \).

  To construct the equivalence on positions,
  let \( s : S \), \( p_0 : \Isolated{P_s} \), \( f : P_s \setminus p_0 \to T \), \( t : T \) and \( q_0 : \Isolated{Q_t} \).
  Extend \( f \) to a map \( \bar{f} \DefEq \Graft_{p_0}(f, t) : P_s \to T \).
  Our goal becomes to construct an equivalence
  \[
      \bigl(
        \sum_{p : P_s} Q_{\bar{f}(p)}
      \bigr) \setminus (p_0 , q_0)
    \simeq
      \bigl(
        \smashoperator{\sum_{p : P_s \setminus p_0}} Q_{f(p)}
      \bigr)
        +
      (Q_t \setminus q_0),
  \]
  which we obtain from \autoref{is-equiv-sigma-remove},
  and by applying the computation rules of grafting to \( \bar{f} : P_s \to T \).
\end{construction}

Note that the above proof essentially factors \( \Chain_{F,G} \) into
\[
  \begin{tikzcd}[column sep=large]
    { \Subst{(\Der{F})}{G} \CTimes \Der{G} }%
      \ar[r, -multimap, "\Chain_{F,G}"]
      \ar[d, multimap-multimap]
      &
    {\Der{(\Subst{F}{G})}}
      \\
    H_0
      \ar[r, -multimap, "\eta"{swap}]
      &
    H_1
      \ar[u, multimap-multimap]
  \end{tikzcd}
\]
in which
\(
  \eta_{\Sh} :
    { \sum_{s : S} \sum_{f : P_s \to T} \sum_{p : \Isolated{P_s}} \Isolated{Q_{fp}} }%
      \to
    { \sum_{s : S} \sum_{f : P_s \to T} \Isolated{\bigl( \sum_{p : P_s} Q_{fp} \bigr)} }
\)
applies \( \SigmaIsolate_{P_s,Q_{f(\Blank)}} \).
We will make this factorization explicit in the derivation of the chain rule for parametrized containers (\autoref{binary-chain-rule}).
We say that the chain rule for \( F \) and \( G \) is \emph{strong} if \( \Chain_{F,G} \) is an equivalence of containers.
Using any of the equivalent properties listed in \autoref{is-equiv-sigma-isolate-iff-isolated-pair},
we can express when exactly we have a strong chain rule.
Note that \( \SigmaIsolate \) is always an embedding (cf.~\autoref{is-embedding-sigma-isolate}),
hence we can think of \( \Chain_{F,G} \) as embedding \( \Subst{(\Der{F})}{G} \CTimes \Der{G} \)
as a subcontainer inside \( \Der(\Subst{F}{G}) \):
\begin{definition}
  A morphism \( f : \Cart{F}{G} \) is an embedding of containers if \( f_{\Sh} : F_{\Sh} \to G_{\Sh} \) is an embedding.%
  \footnote{
    Embeddings are monomorphisms in \( \ContCart \):
    if \( f : \Cart{F}{G} \) is an embedding, then \( f_{\Sh} \) is left-cancellable
    and each \( f_{\Ps}(s) \) is right-cancellable (since it is an equivalence).
    Thus, by extensionality~(\autoref{container-morphism-ext}),
    \( f \circ g = f \circ h \) implies \( g = h \) for parallel \( g, h : \Cart{H}{F} \).
  }
\end{definition}
\begin{proposition}\label{is-embedding-chain-rule}
  For all \( F, G : \Cont \), \( \Chain_{F,G} : \Cart{ \Subst{(\Der{F})}{G} \CTimes \Der{G} }{\Der{(\Subst{F}{G})}} \) is an embedding.
  \begin{proof}
    By \autoref{is-embedding-sigma-isolate}, \( \SigmaIsolate \) is an embedding.
  \end{proof}
\end{proposition}

This strengthens to an equivalence of containers depending on \( \SigmaIsolate \) alone:
\begin{proposition}\label{strong-chain-rule-iff-is-equiv-sigma-isolate}
  Let \( F = (\MkCont{S}{P}) \) and \( G = (\MkCont{T}{Q}) \).
  The following are equivalent propositions:
  \begin{enumerate}
    \item
      \( \Chain_{F,G} \) is an equivalence of containers
    \item
      \( \SigmaIsolate_{P_s,Q_{f(\Blank)}} \) is an equivalence for all \( s : S \) and \( f : P_s \to T \)
  \end{enumerate}
  \begin{proof}
    By inspection of the definition of \( \Chain \) in terms of \( \SigmaIsolate \).
  \end{proof}
\end{proposition}

As expected by \cite[{Proposition~6.6}]{AbbottEtAl2005DataDifferentiating},
the chain rule is strong for \emph{discrete} containers:
\begin{theorem}\label{discrete-strong-chain-rule}
  For discrete containers \( F, G \), \( \Chain_{F,G} \) is an equivalence.
  In particular, it is an isomorphism in the 1-category of set-truncated containers.
  \begin{proof}
    Positions of \( F \) and \( G \) are discrete,
    so by \autoref{discrete-is-equiv-sigma-isolated}
    \( \SigmaIsolate \) is an equivalence.
    By \autoref{strong-chain-rule-iff-is-equiv-sigma-isolate}, \( \Chain_{F,G} \) is an equivalence of containers.
  \end{proof}
\end{theorem}

Secondly, we conclude that globally having a strong chain rule is impossible in the presence of provably non-discrete types:
\begin{theorem}\label{globally-discrete-iff-strong-chain-rule}
  The following are equivalent propositions:
  \begin{enumerate}
    \item \emph{Every} type is discrete.
    \item For all containers \( F \) and \( G\), \( \Chain_{F,G} \) is an equivalence.
  \end{enumerate}
  \begin{proof}
    If every type is discrete, then so is every container, hence the chain rule is always an equivalence by \autoref{discrete-strong-chain-rule}.
    In the other direction,
    use \autoref{discrete-iff-is-equiv-singl-isolate} to show that any given type \( A \) is discrete
    ---
    that is, given some \( a_0 : A \), prove that \( \SigmaIsolate_{A, {a_0 = \Blank}} \) is an equivalence.
    We do so by applying \autoref{strong-chain-rule-iff-is-equiv-sigma-isolate} to containers
    \( F \DefEq (\MkCont{1}{A}) \) and \( G \DefEq (\MkCont{(a : A)}{a_0 = a}) \).
  \end{proof}
\end{theorem}

As a consequence, globally assuming a strong chain rule is inconsistent in the presence of types of higher truncation level:
The circle \( S^1 \) is provably not discrete (if it were, it would be a set!),
hence
\(
  \neg ( \prod_{F, G : \Cont} \Op{isEquiv}(\Chain_{F,G}) )
\).

If instead we restrict ourselves to the world of sets,
then we can conclude that a globally strong chain rule exists if and only if arbitrary equalities are decidable:
\begin{corollary}
  The following are equivalent:
  \begin{enumerate}
    \item Every set is discrete.
    \item In the 1-category of set-truncated containers, \( \Chain_{F,G} \) is an isomorphism
      for all pairs of containers \( F \) and \( G \).
  \end{enumerate}
  \begin{proof}
    In the 1-category \( \ContCart_{0,0} \), the types of equivalences and isomorphisms of containers are equivalent.
    The claim follows by inspection of the proof of \autoref{globally-discrete-iff-strong-chain-rule},
    and ensuring that the same argument applies even when all types involved are sets.
  \end{proof}
\end{corollary}
This is of course still a very classical assumption, even if it is not inconsistent:
In an impredicative setting, where \( \Prop \simeq \Omega \) for some small \( \Omega : \Type \),
either condition implies the law of the excluded middle:
\( \Prop \) is a set, so \( \Omega \) must be as well.
If \( \Omega \) were discrete, \( P = \top \) would be decidable for any \( P : \Prop \).

We have seen that our definition of derivative behaves nicely even in the presence of types which are not discrete,
in that we retain the ways in which it interacts with sums and products (\autoref{sum-product-rule}).
Its interaction with substitution, however, is more subtle.
While we do obtain a chain rule, it is now \emph{directed} or \emph{lax} (\autoref{lax-chain-rule}),
and whether we can strengthen it to an equivalence depends on the pair of containers involved (\autoref{strong-chain-rule-iff-is-equiv-sigma-isolate}).
Indeed, assuming the latter for any pair of containers is inconsistent in the presence of higher inductive types such as the circle \( S^1 \),
or equivalent to a non-constructive principle if restricted to sets.

\section{Derivatives of Fixpoints}\label{fixed-points}

Containers let us model inductive data types as fixpoints to substitution \( \Subst{\Blank}{\Blank} \):
if \( F \) describes the branching of an inductive data type, then there is a corresponding
container \( \mu F \) such that \( \CartIso{\Subst{F}{\mu F}}{\mu F} \).
This lets us turn informal recursive specifications such as \( \Op{List}(X) = 1 + X \times \Op{List}(X) \)
into a precise description of \( \Op{List} \) as a container.
\Citeauthor{AbbottEtAl2005DataDifferentiating} prove that the derivative of a fixpoint container
is itself a fixpoint: \( \Der{(\mu F)} \) is equivalent to \( \mu F' \) for some \( F' \) obtained from \( F \).
To do so, they recognize that each \( \mu F \) essentially behaves like repeated substitution \( \Subst{F}{\Subst{F}{\Subst{F}{\ldots}}} \),
and infer a fixpoint rule from the chain rule.

Our goal is to derive a similar fixpoint rule for our generalized derivative,
ideally by deriving it from the chain rule as well.
This raises the question: if in general the chain rule is lax,
is it ever possible to have a strong fixpoint rule?
It appears that the answer to this question is \enquote{it depends}:
we are able to show (cf.~\autoref{strong-mu-rule-iff-strong-chain-rule})
that the fixpoint rule for container \( F \) is strong if and only if the chain rule
between \( F \) and \( \mu F \) is.

\subsection{Parametrized Containers}

Before we define and reason about an operation \( F \mapsto \mu F \) constructing fixpoints,
we have to recall how types that depend on a number of \emph{type parameters} are modeled by containers.
Ordinary containers express dependency on a single parameter,
but what if we would like to encode the type \enquote{binary trees with \( X \)-labeled nodes and \( Y \)-labeled leaves}?
To encode such parametrized types, we have to index positions by a type representing those \emph{parameters}:
\begin{definition}[note={Parametrized containers}]
  Let \( I : \Type \). The type of \emph{\( I \)-parametrized containers} is
  \(
    \ICont{I} \DefEq \sum_{S : \Type} (I \to S \to \Type)
  \).
\end{definition}
Ordinary containers correspond to unary containers, \( \ICont{1} \).
Each \( i : I \) corresponds to a parameter of the type that a container describes.
Consequently, the \emph{extent} of a parametrized container takes \( I \)-indexed families of types to types,
\begin{align*}
  &\Ext{\MkCont{S}{P}} : (I \to \Type) \to \Type \\
  &\Ext{\MkCont{S}{P}}(\vec{X}) \DefEq \sum_{s : S} \prod_{i : I} P_i(s) \to \vec{X}_i
\end{align*}

Linear morphisms between containers \(F , G : \ICont{I} \) are defined by ranging over all indices \( I \):
\[
  \Cart{F}{G} \DefEq \smashoperator[l]{\sum_{f : F_{\Sh} \to G_{\Sh}}} \prod_{i : I, s : F_{\Sh}} G_{\Ps}(i, f(s)) \simeq F_{\Ps}(i, s)
\]
Together, parametrized containers again assemble into a wild category, \( \IContCart{I} \).
Each container in \( I \) parameters is also one in \( I + 1 \) parameters:
For \( F : \ICont{I} \), denote by \( \Wk{F} : \ICont{I+1} \) the inclusion given by
\[
  \Wk{F}_{\Ps}(\Just{i}) \DefEq F_{\Ps}(i), \quad \Wk{F}_{\Ps}(\Nothing) \DefEq 0
\]
To aid readability, we denote a container \( (\MkCont{S}{\bar{P}}) \) with \( (I+1) \) parameters by \( (\MkCont{S}{\vec{P}, P}) \),
where \( \vec{P}_i \DefEq \bar{P}_{\Just(i)} \) and \( P \DefEq \bar{P}_{\Nothing} \).
In particular, \( \Wk{(\MkCont{S}{P})} = (\MkCont{S}{P,0}) \).
For \( i : I \), the \( i \)th projection container is \( \pi_i \DefEq (\MkCont{1}{\lambda j\,\Blank.\,{i = j}}) : \ICont{I} \).
If \( i \) is isolated, then \( i = j \) is a decidable proposition for any \( j : I \);
in this case \( \pi_i \) is equivalent to a container whose \( i \)th type of positions is \( 1 \),
and \( 0 \) for any other direction.
As expected, parametrized containers are closed under constants \( \CConst(A) \), sums \( (\CPlus) \) and products \( (\CTimes) \).
Substitution generalizes to an operation on containers in \( I + 1 \) parameters;
it places a container inside the positions of the parameter \( \Nothing : I + 1 \):
\begin{definition}
    Substitution is an operation \( \Subst{\Blank}{\Blank} : \ICont{I+1} \to \ICont{I} \to \ICont{I} \),
    defined as follows:
    \begin{align*}
      \Subst{(\MkCont{S}{\vec{P}, P})}{(\MkCont{T}{Q})}_{\Sh} &\DefEq \sum_{s : S} P(s) \to T
        \\
      \Subst{(\MkCont{S}{\vec{P}, P})}{(\MkCont{T}{Q})}_{\Ps} &\DefEq \lambda i\, (s, f).\, \vec{P}_i(s) + \smashoperator{\sum_{p : P(s)}} Q_i(f_p)
        \qedhere
    \end{align*}
\end{definition}
Similarly, we can lift derivatives to parametrized containers.
We can take a derivative with respect to any parameter \( i : I \), as long as equality with \( i \) is decidable:
\begin{definition}
  Let \( F \JudgeEq (\MkCont{S}{P}) : \ICont{I} \) and \( i : \Isolated{I} \).
  The \emph{derivative in parameter \( i \)} is the container \( {\Der_i}F : \ICont{I} \) defined as follows:
  \[
    {\Der_i(F)} \DefEq \left(
      \MkCont%
        {\sum_{s : S} \Isolated{P_i(s)}}%
        {\lambda (j , s , p)\Where
          \begin{cases}
            P_i(s) \setminus p & \text{if } i = j \\
            P_j(s) & \text{otherwise}
          \end{cases}
        }%
    \right)
  \]
\end{definition}

In order to reduce visual clutter, we investigate derivatives of containers in two parameters,
that is, those parametrized by the type \( 2 = 1 + 1 \).
However, our arguments apply to containers of \( (I + 1) \) parameters in general.
In particular, we write \( \Der_0 \) and \( \Der_1 \) for the two possible derivatives of two-parameter containers,
that is \( \Der_{\Inl(\bullet)} \) and \( \Der_{\Inr(\bullet)} \).
In the unary case, we omit the subscript and simply write \( \Der \) for \( \Der_{\bullet} \).

Derivatives of parametrized containers satisfy the same basic laws for constants, sums and products as in the unary case.
Derivatives of composites follow a lax chain rule, which now accounts for multiple indices.
As promised in the discussion of \autoref{lax-chain-rule},
its construction factors into smaller steps that make it obvious why, in general, it is not invertible:
\begin{problem}[note={Lax chain rule for two-parameter containers}]\label{binary-chain-rule}
  Let \( F : \ICont{2} \) and \( G : \ICont{1} \).
  Define a linear morphism
  \[
    \Chain_{F,G}
      :
    \Cart%
      {(\Der_0{F})[G] \CPlus \bigl( (\Der_1{F})[G] \CTimes \Der{G} \bigr)}%
      {\Der{(F[G])}}
  \]
  \begin{construction}
    \renewcommand*{\L}{\Op{L}}
    \newcommand*{\R}{\Op{R}}
    Let \( F \JudgeEq (\MkCont{S}{P}) \), \( G \JudgeEq (\MkCont{T}{Q}) \).
    We define auxiliary containers \( H_1, H_2 \) and factor the morphism through a pair of equivalences as follows:
    \[
      \begin{tikzcd}[column sep=large]
        {\mathllap{\L \DefEq {}} {(\Der_0{F})[G] \CPlus \bigl( (\Der_1{F})[G] \CTimes \Der{G} \bigr)}}
          \ar[d, multimap-multimap]
          \ar[r, -multimap, dashed, "\Chain_{F,G}"]
          &
        {\Der{(F[G])} \mathrlap{{} \eqcolon \R}}
          \ar[d, multimap-multimap]
          \\
        H_1
          \ar[r, -multimap, "\eta"{swap}]
          &
        H_2
      \end{tikzcd}
    \]
    Following the argument in \autoref{lax-chain-rule}, we apply \autoref{graft-equiv} to graft a function typed \( P_1(s) \to T \) in \eqref{binary-chain-rule-shape-equiv},
    and see that the shapes of \( \L \) are equivalent to
    \begin{align}
      \L_{\Sh}
        &\simeq
          \sum_{s : S} \left(\Isolated{P_0(s)} \times (P_1(s) \to T)\right) + \smashoperator[l]{\sum_{p : \Isolated{P_1(s)}}} \sum_{(f, t) : \Highlight{(P_1(s) \setminus p \to T) \times T}} {\Isolated{Q(t)}}
          \notag
          \\
        &\simeq
          \sum_{s : S} \left(\Isolated{P_0(s)} \times (P_1(s) \to T)\right) + \smashoperator[l]{\sum_{p : \Isolated{P_1(s)}}} \sum_{f : \Highlight{P_1(s) \to T}} {\Isolated{Q(f_p)}}
          \label{binary-chain-rule-shape-equiv}%
          \\
        &\simeq
          \sum_{s : S} \smashoperator[r]{\sum_{f : P_1(s) \to T}} \,\bigl(\Isolated{P_0(s)} + \sum\nolimits_{p : \Isolated{P_1(s)}} \Isolated{Q(f_p)}\bigr)
          \eqcolon U_1
          \notag
    \end{align}
    Denote this equivalence by \( f_1 : {\L}_{\Sh} \simeq U_1 \), and let \( H_1 \DefEq (\MkCont{U_1}{{\L}_{\Ps} \circ f_1^{\Inv}}) \).
    This defines the equivalence \( \CartIso{\L}{H_1} \) to the left.

    On the other side we obtain an equivalence \( f_2 : U_2 \simeq {\R}_{\Sh} \)
    by distributing \( \Isolated{(\Blank)} \) over binary sums:
    \begin{align*}
      U_2
      &\DefEq
      \sum_{s : S} \sum_{f : P_1(s) \to T}
        \Isolated{\Highlight{P_0(s)}} + \Isolated{\Bigl( \HighlightAlt{ \sum\nolimits_{p : P_1(s)} Q(f_p) } \Bigr)}
        \\
      &\simeq
      \sum_{s : S} \sum_{f : P_1(s) \to T}
        \Isolated{\bigl( \Highlight{P_0(s)} + \HighlightAlt{ \sum\nolimits_{p : {P_1(s)}} Q(f_p) } \bigr)}
        \\
      &\simeq
        {\R}_{\Sh}
    \end{align*}
    Let \( H_2 \DefEq (\MkCont{U_2}{{\R}_{\Ps} \circ f_2}) \),
    which yields the equivalence \( \CartIso{H_2}{\R} \) to the right.

    Let us now define \( \eta : H_1 \multimap H_2 \).
    As in \autoref{lax-chain-rule}, define the shape map \( \eta_{\Sh} : U_1 \to U_2 \) using \( \SigmaIsolate_{P_1(s),Q(f(\Blank))} \).
    On positions, the equivalence
    \(
      \eta_{\Ps}(u) : H_2^{\Ps}(\eta_{\Sh}(u)) \simeq H_1^{\Ps}(u)
    \)
    is defined by cases, depending on which side of the sum \( u : U_1 \) falls.
    Let \( s : S \), \( f : P_1(s) \to T \).
    In case of \( u \JudgeEq (s, f, \Inl{p_0}) \) for \( p_0 : \Isolated{P_0(s)} \), our goal is to give
    \[
      ( \Isolated{P_0(s)} + B ) \setminus \Inl(p_0)
        \simeq
      ( \Isolated{P_0(s)} \setminus p_0 ) + B
    \]
    for \( B \DefEq {\textstyle \sum_{p : P_1(s)} Q(f_p) } \), which is an instance of \autoref{sum-remove-equiv}.

    When \( u \JudgeEq (s, f, \Inr(p_1, q)) \) for some \( p_1 : \Isolated{P_1(s)} \) and \( q : \Isolated{Q(f_p)} \),
    we rewrite the type of positions as follows:
    \begin{align}
      H_2^{\Ps}(\eta_{\Sh}(s , f , \Inr(p_1, q)))
        &\mathrel{\JudgeEq}
          \bigl( P_0(s) + \sum\nolimits_{p : P_1(s)} Q(f_p) \bigr) \setminus \Inr(p_1, q)
          \notag
          \\
        &\simeq
          P_0(s) + \Bigl( \sum\nolimits_{p : P_1(s)} Q(f_p) \Bigr) \setminus (p_1, q)
          \label{binary-chain-rule-pos-equiv-sum-minus}
          \\
        &\simeq
          P_0(s) + \sum\nolimits_{p : P_1(s) \setminus p} Q(f_p) + (Q(f_{p_1}) \setminus q)
          \label{binary-chain-rule-pos-equiv-sigma-minus}
          \\
        &\mathrel{\JudgeEq}
          H_1^{\Ps}(s, f, \Inr{(p_1, q)})
          \notag
    \end{align}
    In \eqref{binary-chain-rule-pos-equiv-sum-minus}, we move the pair \( (p_1, q) \) to the right of the sum (\autoref{sum-remove-equiv}).
    For \eqref{binary-chain-rule-pos-equiv-sigma-minus},
    we split the \( \Sigma \)-type by applying \autoref{is-equiv-sigma-remove}.
    This is justified since both \( p_1 \) and \( q \) are isolated points,
    and together, the pair \( (p_1, q) \) is isolated in \( \sum_{p : P_1(s)} Q(f_p) \) by \autoref{is-isolated-pair}.
  \end{construction}
\end{problem}

This factorization makes it obvious that the chain rule for two-parameter containers is again an embedding of containers,
and strong whenever isolated points distribute over dependent sums:
\begin{proposition}\label{is-embedding-binary-chain-rule}
  For \( F : \ICont{2} \) and \( G : \ICont{1} \), \( \Chain_{F,G} \) is an embedding. \qed
\end{proposition}
\begin{proposition}\label{strong-binary-chain-rule-iff-is-equiv-sigma-isolate}
  For all \( F \JudgeEq (\MkCont{S}{P}) : \ICont{2} \) and \( G \JudgeEq (\MkCont{T}{Q}) : \ICont{1} \), the following are equivalent:
  \begin{enumerate}
    \item \( \Chain_{F,G} \) is an equivalence of unary containers.
    \item For all \( s : S \) and \( f : P_1(s) \to T \), \( \SigmaIsolate_{P_1(s), Q(f(\Blank))} \) is an equivalence.
  \end{enumerate}
  \begin{proof}
    The morphism \( \Chain_{F,G} \) is an equivalence if and only if \( \eta_{\Sh} \) in the above construction is an equivalence of types.
    This in turn is exactly the case when the second condition holds.
  \end{proof}
\end{proposition}

Like in \autoref{discrete-strong-chain-rule}, this is the case for discrete containers:
\begin{proposition}\label{discrete-strong-binary-chain-rule}
  If \( F : \ICont{2} \) and \( G : \ICont{1} \) are discrete, then \( \Chain_{F,G} \) is an equivalence. \qed
\end{proposition}

\subsection{Fixpoints of Containers}\label{fixed-point-containers}

To show how derivatives of untruncated containers interact with fixpoints,
we first have to give a suitable specification of what we mean by such a \enquote{fixpoint}.
In the traditional understanding of parametric polymorphism as functors and natural transformations {\`a la}~\cite{BainbridgeEtAl1990Functorialpolymorphism},
an inductive type is a (least) solution to a fixpoint problem \( Y = F(\vec{X}, Y) \)
for some functor \( F \) depending on some number of (type) parameters \( \vec{X} \).
This perspective is easily adapted when reasoning in terms of containers.
For any \emph{set-truncated} container \( {F : \ICont{I+1}} \),
the fixpoint problem \( Y = \Ext{F}(\vec{X}, Y) \) always has a unique least solution:
the endofunctor \( \Ext{F}(\vec{X}, \Blank) : [ \Set, \Set ] \) admits an initial algebra,
naturally for all parameters \( {\vec{X} : I \to \Set} \)~\cite[Proposition~5.3]{AbbottEtAl2005ContainersConstructingstrictly}.
This initial algebra is itself the extent of some container \( \mu F : \ICont{I} \).
The embedding of set-truncated containers into \( [\Set^I, \Set] \) via \( \Ext{\Blank} \) is full and faithful,
so \( \mu F \) represents a \emph{unique} solution to the fixpoint problem \( Y = \Ext{F}(\vec{X}, Y) \).

Recently,
\citeauthor{DamatoEtAl2025FormalisingInductiveCoinductive} have investigated this in a univalent setting~\cite{DamatoEtAl2025FormalisingInductiveCoinductive}.
They show, without assuming \( \Op{UIP} \), that fixpoints exist at the level of container \emph{functors}:
for any container \( F : \ICont{I + 1} \) and all parameters \( \vec{X} : I \to \Type \),
the wild functor \( \Ext{ F }(\vec{X}, \Blank) : [\Type, \Type] \) admits an initial algebra
whose carrier is given by \( \Ext{ \mu F }\vec{X} \).
Unfortunately, we cannot (yet) internally prove that \( \Ext{\Blank} \) is a suitable embedding of the \( (\infty, 1) \)-category of containers
into \( \Type \)-valued endofunctors, preventing us from showing that \( \mu F \) is unique among containers.
A candidate proof by \citeauthor{Damato2023RevisitingContainersCubical} based on the Yoneda lemma~\cite{Damato2023RevisitingContainersCubical}
should be amenable to generalization to \( (\infty, 1) \)-categories,
but alas, it is unclear how to define those internal to HoTT.

We instead observe that in a univalent setting, it is easier to follow \citeauthor{Abbott2003CategoriesContainers}'s original conception of container fixpoint~\cite[Definition~5.1.6]{Abbott2003CategoriesContainers} directly:
We prove that substitution \( \Subst{F}{\Blank} \) into a fixed \( F : \ICont{I+1} \) itself is a wild endofunctor on containers,
and that the ordinary fixpoint container \( \mu F \) is the carrier of its initial algebra.
The construction of the container \( \mu F \) itself is well known,
but unlike previous work~\cite{Abbott2003CategoriesContainers,AbbottEtAl2005ContainersConstructingstrictly,AltenkirchEtAl2015Indexedcontainers,DamatoEtAl2025FormalisingInductiveCoinductive},
we do not reason about the associated polynomial functor \( \Ext{\mu F} \) to prove that \( \mu F \) is a fixpoint.
Thus, our proofs avoid the intricacies of \( \Ext{\Blank} \) as an \( (\infty, 1) \)-functor imposed on us in a univalent setting.
At times, this even simplifies proofs --- at least when reasoning about morphisms, we never have to verify naturality conditions.
Concretely, we first prove that \( \Subst{F}{\Blank} \)-algebras form a wild category
(which is not the case for arbitrary wild endofunctors),
and then construct an initial object in this wild category of algebras.

\begin{problem}
  Given \( F : \ICont{I+1} \), extend \( \Subst{F}{\Blank} \) to a wild endofunctor on \( \IContCart{I} \)\kern-.2em.
\end{problem}
\begin{construction}
  Let \( F \JudgeEq (\MkCont{S}{\vec{P}, P}) \).
  Given \( h : (\MkCont{T}{Q}) \to (\MkCont{U}{R}) \),
  define the shape map by postcomposition,
  \[
    \Subst{F}{h}_{\Sh} \DefEq \Sigma(\Op{id}_S, h_{\Sh} \circ \Blank) : {\textstyle \sum_{s : S} (P_s \to T) \to \sum_{s : S} (P_s \to U) }.
  \]
  On positions, given \( s : S \) and \( f : P_s \to T \), apply \( h_{\Ps}(i, f_p) : R_i(h_{\Sh}(f_p)) \simeq Q_i(f_p) \)
  to the second component of the \( \Sigma \) of
  \[
    \Subst{F}{h}_{\Ps}(i, (s, f)) : {\textstyle
      \bigl(\vec{P}_i(s) + \sum_{p \in P_s} R_i(h_{\Sh}(f_p)) \bigr)
        \simeq
      \bigl(\vec{P}_i(s) + \sum_{p \in P_s} Q_i(f_p)\bigr)
      }
  \]
  By extensionality of container morphisms, this action preserves identities and composites.
\end{construction}
\begin{problem}[note={\( \Subst{F}{\Blank} \)-algebras}]
  Given \( F : \ICont{I+1} \), define the structure of a wild category \( \Alg{\Subst{F}{\Blank}} \).
  Its objects are of type \( \sum_{X : \ICont{I}} \Cart{\Subst{F}{X}}{X} \),
  and a morphism is of type
  \[
    \Alg{\Subst{F}{\Blank}}((X, f), (Y , g)) \DefEq \sum_{h : \Cart{X}{Y}} (h \circ f = g \circ \Subst{F}{h}),
  \]
  i.e.\@ a morphism \( h : \Cart{X}{Y} \) together with a path \( h_\square \) ensuring that the square
  \[
    \begin{tikzcd}
      \Subst{F}{X} \ar[r, -multimap, "\Subst{F}{h}", ""{swap,name=Ff}] & \Subst{F}{Y} \\
      X \ar[r, -multimap, "h"{swap}, ""{name=f}] & Y
      \ar[from=1-1, to=2-1, -multimap, "f"{swap}]
      \ar[from=1-2, to=2-2, -multimap, "g"]
      \ar[from=Ff, to=f, phantom, "{\scriptstyle h_\square}"]
    \end{tikzcd}
  \]
  commutes.
\end{problem}
\begin{construction}
  Identities and composite morphisms are defined as expected.
  The filler of the commutative square in a composite \( (h, h_\square) ; (k, k_\square) \) is
  obtained by path composition of \( h_\square \) and \( k_\square \).
  However, unlike in 1-categories, unitality and associativity of composition are not automatically inherited from the base category \( \ICont{I} \),
  since we have to ensure that there are higher paths between composite squares, such as
  \[
    \begin{tikzcd}
      \Subst{F}{X} \ar[r, -multimap, "\Subst{F}{h}", ""{swap,name=Ff}]
        & \Subst{F}{Y} \ar[r, -multimap, "\Subst{F}{\id}", ""{swap,name=Fid}]
        & \Subst{F}{Y}  \\
      X \ar[r, -multimap, "h"{swap}, ""{name=f}]
        & Y \ar[r, -multimap, "\id"{swap}, ""{name=id}]
        & Y
      \ar[from=1-1, to=2-1, -multimap, "f"{swap}]
      \ar[from=1-2, to=2-2, -multimap, "g"]
      \ar[from=1-3, to=2-3, -multimap, "g"]
      \ar[from=Ff, to=f, phantom, "{\scriptstyle h_\square}"]
      \ar[from=Fid, to=id, phantom, "{\scriptstyle \id_\square}"]
    \end{tikzcd}
    =
    \begin{tikzcd}
      \Subst{F}{X} \ar[r, -multimap, "\Subst{F}{h}", ""{swap,name=Ff}] & \Subst{F}{Y} \\
      X \ar[r, -multimap, "h"{swap}, ""{name=f}] & Y
      \ar[from=1-1, to=2-1, -multimap, "f"{swap}]
      \ar[from=1-2, to=2-2, -multimap, "g"]
      \ar[from=Ff, to=f, phantom, "{\scriptstyle f_\square}"]
    \end{tikzcd}
  \]
  These higher paths are constructed by extensionality of container morphisms.
  On shapes, the necessary coherences reduce to showing that path composition is unital and associative.
  The coherences on positions are obtained following the same steps,
  except that they are now displayed over the coherences of shapes.\footnote{%
    We omit the construction of this wild category from the formalization.
    Defining the necessary coherence between commutative squares is challenging for technical reasons;
    a formal proof only obscures the intuition that \( \Subst{F}{\Blank} \) acts coherently also on homotopies of container morphisms.
    At the same time, no further construction depends on these coherences.
  }
\end{construction}

Unlike other concepts, we can express initiality in wild categories coherently as a \emph{property} of an object:
an object \( 0 \) of a wild category \( \mathcal{C} \) is \emph{initial}
if for any other object \( X \), the type of morphisms \( \mathcal{C}(0, X) \) is \emph{contractible}.
Hence we say:
\begin{definition}[note={\cite[Definition~5.1.6]{Abbott2003CategoriesContainers}}]
  A least fixpoint of a container \( F : \ICont{I + 1} \) is an initial \( \Subst{F}{\Blank} \)-algebra.
\end{definition}
Spelled out, a least \( F \)-fixpoint consists of a \emph{carrier} \( M : \ICont{I} \)
and a structure map \( m : \Cart{\Subst{F}{M}}{M} \)
such that for any other algebra \( (G, g) \), the type of morphisms \( \Alg{\Subst{F}{\Blank}}((M, m), (G, g)) \) is contractible.
The latter means that there is a unique map \( r : \Cart{M}{G} \) that commutes with the structure maps \( m \) and \( g \).
As a consequence, the carriers of least fixpoints are equivalent (hence equal) as containers.

We now show, assuming access to \( \W \)-types, that all containers admit least fixpoints in this sense:
Recall that for \( A : \Type \) and \( B : A \to \Type \),
the type \( \W_{A}(B) \) has the single constructor
\begin{equation*}
  {
    \AxiomC{\( a : A \)}
    \AxiomC{\( f : B(a) \to \W_{A}(B) \)}
    \BinaryInfC{\( \Sup(a, f) : \W_{A}(B) \)}
    \DisplayProof
  }
\end{equation*}
Each \( w : \W_{A}(B) \) is a well-founded tree with \( A \)-labeled nodes, and \( B(a) \)-branching subtrees.
A finite path to some node inside \( w \), labeled by a family \( {C : A \to \Type} \),
can be defined as an inductive family \( \WPath_{A,B,C} : \W_{A}(B) \to \Type \) with two constructors, namely
\begin{center}
  \hspace*{\fill}
  {
    \AxiomC{\( c : C(a) \)}
    \UnaryInfC{\( \WTop(c) : \WPath_{A, B, C}(\Sup(a, f)) \)}
    \DisplayProof
  }
  \hfill%
  and
  \hfill%
  {
    \AxiomC{\( b : B(a) \)}
    \AxiomC{\( w : \WPath_{A,B,C}(f(b)) \)}
    \BinaryInfC{\( \WBelow(b, w) : \WPath_{A, B, C}(\Sup(a, f)) \)}
    \DisplayProof
  }
  \hspace*{\fill}
\end{center}
for all \( a : A \) and \( f : B(a) \to \W_{A}(B) \).
Both \( \W \) and \( \WPath \) can be described by unfolding them by one level
---
the constructors define equivalences
\begin{align*}
  \WIn &: \sum_{a : A} (B(a) \to \W_{A}(B)) \simeq \W_{A}(B)
    \\
  \WPathIn_{a, f} &: C(a) + \smashoperator{\sum_{b : B(a)}} \WPath_{A,B,C}(f(b)) \simeq \WPath_{A,B,C}(\Sup(a, f))
\end{align*}
for all \( a : A \) and \( f : B(a) \to \W_{A}(B)\).

We follow \cites[\S~5]{AbbottEtAl2005ContainersConstructingstrictly}[\S~7]{AbbottEtAl2005DataDifferentiating} and
use \( \W \) and \( \WPath \) to define the shapes and positions, respectively, of the least fixpoint container:
\begin{definition}
  Let \( F \JudgeEq (\MkCont{S}{\vec{P},P}) : \ICont{I+1} \), define \( {\mu F} \DefEq (\MkCont{S^\mu}{P^\mu}) : \ICont{I} \):
  \begin{align*}
    S^\mu &\DefEq \W_{S}(P) \\
    P^\mu_i & \DefEq \WPath_{S,P}(\vec{P}_i)
    \qedhere
  \end{align*}
\end{definition}

\begin{example}
  \newcommand*{\Nil}{\Op{nil}}
  \newcommand*{\Cons}{\Op{cons}}
  In the informal specification \( \Op{List}(X) = \mu Y.\, 1 + X \times Y \),
  the expression \( L(X, Y) = 1 + X \times Y \) is transcribed into a container in two parameters, \( L \JudgeEq (\MkCont{S}{X, Y}) : \Cont_2 \).
  It has two shapes, \( S \DefEq \{ \Nil,  \Cons \} \),
  over which there are positions
  \begin{align*}
    X(\Nil) &\DefEq 0 & X(\Cons) &\DefEq 1 \\
    Y(\Nil) &\DefEq 0 & Y(\Cons) &\DefEq 1
  \end{align*}
  Its fixpoint is equivalent to the usual description of lists as a container:
  the shapes \( ({\mu L})_{\Sh} \JudgeEq \W_{S}(Y) \)
  are trees with either unary (\( Y(\Cons) \)) or nullary (\( Y(\Nil) \))
  branching, hence there is a unique tree for each depth \( n : \Nat \).
  An \( X \)-labeled path \( x : \WPath_{S, Y, X}(t) \) to a node in a tree \( t : \W_{S}(Y) \)
  thus corresponds exactly to some number \( k : \Fin{\Op{depth}(t)} \).
  This defines an equivalence \( \mu L \CartEquiv (\MkCont{n : \Nat}{\Fin{n}}) \).
\end{example}

In general, \( \mu F \) is a fixpoint of substitution into \( \Subst{F}{\Blank} \):
\begin{problem}
  Define an equivalence of containers \( \In_F : F[{\mu F}] \CartEquiv {\mu F} \).
  \begin{construction}
    Let \( F \JudgeEq (\MkCont{S}{\vec{P},P}) \).
    Shapes of \( \Subst{F}{\mu F} \) are \( \sum_{s : S} (P(s) \to \W_{S}(P)) \),
    hence \( \WIn \) establishes an equivalence with the shapes of \( \mu F \).
    Similarly, the positions of \( \Subst{F}{\mu F} \) are equivalent to those of \( \mu F \) via \( \WPathIn \).
  \end{construction}
\end{problem}

Together, \( \mu F \) and \( \In_F \) form an \( \Subst{F}{\Blank} \)-algebra.
We define a non-dependent recursion principle for maps out of \( \mu F \),
which yields a morphism into \emph{any} other algebra:
\begin{problem}[note={\( \mu \)-recursion}]\label{mu-rec}
  Define a recursion principle for morphisms out of \( \mu \)-containers.
  That is, for any \( F : \ICont{I+1} \), \( G : \ICont{I} \) and \( \alpha : \Cart{\Subst{F}{G}}{G} \),
  give a morphism
  \[
      \Rec_{F}(\alpha) : \Cart{\mu F}{G}
  \]
  together with a chosen filler
  \begin{equation}\label{mu-rec-comm}
    \begin{tikzcd}[column sep=huge]
      \Subst{F}{\mu F} & \Subst{F}{G} \\
      {\mu F}          & G
      \ar[from=1-1, to=1-2, -multimap, "\Subst{F}{\Rec_F(\alpha)}"]
      \ar[from=1-1, to=2-1, -multimap, "\In_F"{swap}]
      \ar[from=2-1, to=2-2, -multimap, "\Rec_F(\alpha)"{swap}]
      \ar[from=1-2, to=2-2, -multimap, "\alpha"]
    \end{tikzcd}
  \end{equation}
\end{problem}
\begin{construction}
  Let \( F \JudgeEq (\MkCont{S}{\vec{P},P}) \) and \( G \JudgeEq (\MkCont{T}{Q}) \).
  Abbreviate \( \Rec \DefEq \Rec_{F}(\alpha) \), \( \W \DefEq \W_{S}(P) \) and \( \WPath \DefEq \WPath(S,P,\vec{P}_i) \).
  On shapes, \( \Rec \)  is defined by \( \W \)-recursion,
  \begin{align*}
    \Rec_{\Sh} &: \W \to T \\
    \Rec_{\Sh} &\DefEq \lambda(\Sup(s, f)).\, \alpha_{\Sh} (s, \Rec_{\Sh} \circ f)
  \end{align*}
  On positions we define
  \( u : \prod_{w : \W_{S}(P)} Q_i(\Rec_{\Sh}(w)) \to \WPath(w) \) by \( \W \)-induction.
  If \( w \JudgeEq \Sup(s, f) \), then \( u(w) \) is the composite
  \[
    \begin{tikzcd}
      Q_i(\alpha_{\Sh}(s , \Rec_{\Sh} \circ f))
        \ar[r, dashed, "u"]
        \ar[d, "{\alpha_{\Ps}(i, \Rec_{\Sh} \circ f)}"{swap}]
        &
      \WPath(\Sup(s, f))
        \\
      \vec{P}_i(s) + \sum_{p \in P(s)} Q_i(\Rec_{\Sh}(f(p)))
        \ar[r, "u^*"{swap}]
        &
      \vec{P}_i(s) + \sum_{p \in P(s)} \WPath(f(p))
        \ar[u, "\WPathIn"{swap}]
    \end{tikzcd}
  \]
  in which \( u^* \) applies \( u(f(p)) \) recursively in the second component of the \( \Sigma \)-type.
  By induction, all maps in this composite are equivalences, hence it is the map underlying
  the desired equivalence of positions, \( \Rec_{\Ps}(i, w) \).

  To show that \eqref{mu-rec-comm} commutes, it suffices to find a filler for
  \begin{equation}\label{mu-rec-out-filler}
    \begin{tikzcd}[column sep=huge]
      \Subst{F}{\mu F} & \Subst{F}{G} \\
      {\mu F}          & G
      \ar[from=1-1, to=1-2, -multimap, "\Subst{F}{\Rec_F(\alpha)}"]
      \ar[from=1-1, to=2-1, multimap-, "\Out_F"{swap}]
      \ar[from=2-1, to=2-2, -multimap, "\Rec_F(\alpha)"{swap}]
      \ar[from=1-2, to=2-2, -multimap, "\alpha"]
    \end{tikzcd}
  \end{equation}
  as \( \In_F \) is an equivalence with inverse \( \Out_F \).
  This is done by \( \W \)-induction on the shapes of \( \mu F \).
\end{construction}

In fact, this morphism is uniquely determined, hence we get a fixpoint for any container in \( I + 1 \) parameters.
We establish this by proving that \( \Rec_F(\alpha) \) is unique as a coalgebra-to-algebra morphism:
\begin{proposition}\label{mu-rec-coalg-to-alg-unique}
  For all \( F : \ICont{I+1} \), \( G : \ICont{I} \) and \( \alpha : \Cart{\Subst{F}{G}}{G} \),
  the type
  \[
    \sum_{\alpha^* : \Cart{\mu F}{G}} \alpha^* = \Out_F ; \Subst{F}{\alpha^*} ; \alpha
  \]
  is contractible with center \( (\Rec_F(\alpha), \Rec_F(\alpha)_\square) \), where
  \( \Rec_F(\alpha)_\square \) is the filler of \eqref{mu-rec-out-filler}.
  \begin{proof}
    Let \( \alpha^* : \Cart{\mu F}{G} \) and \( \alpha^*_\square : \alpha^* = \Out_F ; \Subst{F}{\alpha^*} ; \alpha \).
    We need to give a path \( p : \Rec_F(\alpha) = \alpha^* \),
    and over it a dependent path \( p_\square :  \Rec_F(\alpha)_\square =_p \alpha^*_\square \) connecting the commuting squares
    \[
      \begin{tikzcd}[column sep=huge]
        \Subst{F}{\mu F} & \Subst{F}{G} \\
        {\mu F}          & G
        \ar[from=1-1, to=1-2, -multimap, "\Subst{F}{\Rec_F(\alpha)}"]
        \ar[from=1-1, to=2-1, multimap-, "\Out_F"{swap}]
        \ar[from=2-1, to=2-2, -multimap, "\Rec_F(\alpha)"{swap}]
        \ar[from=1-2, to=2-2, -multimap, "\alpha"]
      \end{tikzcd}
      \qquad\text{and}\qquad
      \begin{tikzcd}[column sep=huge]
        \Subst{F}{\mu F} & \Subst{F}{G} \\
        {\mu F}          & G
        \ar[from=1-1, to=1-2, -multimap, "\Subst{F}{\alpha^*}"]
        \ar[from=1-1, to=2-1, multimap-, "\Out_F"{swap}]
        \ar[from=2-1, to=2-2, -multimap, "\alpha^*"{swap}]
        \ar[from=1-2, to=2-2, -multimap, "\alpha"]
      \end{tikzcd}
    \]
    Let us sketch the proof.
    By extensionality of morphisms (\autoref{container-morphism-ext}), the path \( p \) can be given as a \( {\mu F}_{\!\Sh} \)-indexed family of paths.
    But \( {\mu F}_{\!\Sh} \) is a \( \W \)-type,
    hence we define \( p \) by \( \W \)-induction as a composite of two paths, \( p_1 \) and \( p_2 \), using the path \( \alpha^*_{\square} \).
    The composite of these two paths is unique up to a unique dependent path, and from this we build the filler connecting the two squares above.
    In our mechanized proof, this construction makes extensive use of Cubical Agda's primitive support for dependent path types and cube filling operations.
    This allows us to leave \texttt{transport} hell, and instead settle in \texttt{PathP}  purgatory.
  \end{proof}
\end{proposition}

From this, uniqueness as an algebra follows immediately,
assuming that the relevant commutative squares have equivalent types of fillers.
Unlike for set-truncated containers (where commutativity of both squares is a proposition)
these fillers are data, and we have to ensure that we preserve their structure.
\begin{theorem}\label{mu-initial-algebra}
  For all \( F : \ICont{I + 1}\), the pair \( (\mu F, \In_F) \) is an initial \( \Subst{F}{\Blank} \)-algebra,
  hence a least fixpoint of \( F \).
  \begin{proof}
    Let \( (G, \alpha) \) be a second \( \Subst{F}{\Blank} \)-algebra.
    The claim follows from \autoref{mu-rec-coalg-to-alg-unique} if the respective types of square-fillers are equivalent.
    We have
    \[
      (\alpha^* = \Out_F ; \Subst{F}{\alpha^*} ; \alpha)
        \overset{(?)}{\simeq}
      (\In_F ; \alpha^* = \In_F ; \Out_F ; \Subst{F}{\alpha^*} ; \alpha)
        \simeq
      (\In_F ; \alpha^* = \Subst{F}{\alpha^*} ; \alpha)
    \]
    in which \( (?) \) is an application of \autoref{container-precomp-equiv}:
    precomposition with \( \In_F \) is an equivalence of types
    \( \mu F \to G \) and \( \Subst{F}{\mu F} \to G \).
  \end{proof}
\end{theorem}

With this we have shown that the construction of least fixpoints of containers
{\`a} la \cite{AltenkirchEtAl2015Indexedcontainers} is valid in a univalent setting where \( \Op{UIP} \) is unavailable.
We succeeded because we aligned our proof obligations with the only primitive at hand:
\( \mu F \) is defined in terms of \( \W \)-types,
so any property has to ultimately be established by \( \W \)-induction.
Instead of proving initiality directly like \cite[Proposition~5.2]{AltenkirchEtAl2015Indexedcontainers} does,
we go through \autoref{mu-rec-coalg-to-alg-unique} to show that \( \Rec \) is unique as a coalgebra-to-algebra morphism.
Since all proof obligations (namely \( p \) and \( p_\square \) therein) are ultimately \( \W \)-indexed families,
this can be shown by \( \W \)-induction alone,
without any contentious transport that might require applications of \( \Op{UIP} \).
Assuming \( \Op{UIP} \), the proof of \autoref{mu-initial-algebra} becomes trivial:
the filler-types are inhabited propositions, so trivially equivalent.
Our proof depends on univalence in the form of \autoref{container-precomp-equiv},
and ultimately the fact that equivalence of containers is an identity system.

Introducing a cut in this argument may seem inconsequential,
but we believe that it nicely separates the computational aspect
from the specification
---
on one hand \( \Rec \) is a unique fold-like operation (\autoref{mu-rec-coalg-to-alg-unique}),
on the other \( \mu F \) is the minimal solution to a problem (\autoref{mu-initial-algebra}).

For conciseness, we have stated the results of this section only for the subcategory of linear morphisms.
In general, \( \Subst{F}{\Blank} \) is a wild endofunctor on \( \ICont{I} \),
and \( \mu F \) its initial algebra.
We refer the reader to \S~4.3 of the first-named author's thesis~\cite{Joram2026DatatypesSymmetriesHomotopy}
for a treatment that separates the linear and non-linear case.

In the future, we would like to apply the dual argument to greatest \( \Subst{F}{\Blank} \)-fixpoints,
where unfolding \( \Op{corec} : G \to \nu F \) gives rise to a terminal coalgebra on \( \nu F \).
The shapes of this fixpoint are an \emph{\( \Op{M} \)-type},
whose existence is derivable both in a setting with \( \Op{UIP} \)~\cite{AltenkirchEtAl2015Indexedcontainers},
and assuming univalence~\cite{AhrensEtAl2015NonWellfoundedTrees}.
In particular, applications of \( \W \)-induction in \autoref{mu-rec-coalg-to-alg-unique}
should translate into applications of the coinduction proof principle of \( \Op{M} \)-types.
This would give us a unique algebra-to-coalgebra morphism,
which would yield a greatest \( \Subst{F}{\Blank} \)-fixpoint.

\subsection{The \texorpdfstring{\( \mu \)-}{Fixpoint }Rule}

Having convinced ourselves that fixpoints do exist even for untruncated containers, we can now characterize their derivatives.
Let \( F \JudgeEq (\MkCont{S}{P_0, P_1}) : \ICont{2} \).
If we think of \( \mu F \) as an \( F \)-branching tree, then \( \Der{\mu F} \) represents a type of trees with a hole.
Such a hole must be either in the position of a leaf or occur recursively in a branch,
which are given by the positions \( P_0 \) and \( P_1 \).
Hence, the derivative of \( \mu F \) is again tree-shaped,
with branching given by \( \Der_0{F} \) and \( \Der_1{F} \):
If \( F \) is discrete, there is an equivalence
\(
  \CartIso { \mu F' }{ \Der(\mu F) }
\)
in which
\(
  F' \DefEq
    { \Wk{\Der_0{F}[ {\mu F} ]} }
      \CPlus
    (
      \Wk{\Der_1{F}[ {\mu F} ]}
        \CTimes
      \Proj{1}
    )
\).
Proving that this is the case even for untruncated containers is not entirely straightforward.
There is however hope to recover at least a morphism out of \( \mu F' \) into \( \Der{\mu F} \) by recursion:

\begin{problem}[note={Lax \( \mu \)-rule}]\label{mu-rule}
  For \( F : \IContCart{2} \), define by recursion a linear morphism
  \[
    \MuRule_F : \Cart{ \mu F' }{ \Der(\mu F) }
  \]
  where
  \(
    F' \DefEq
      { \Wk{\Der_0{F}[ {\mu F} ]} }
        \CPlus
      (
        \Wk{\Der_1{F}[ {\mu F} ]}
          \CTimes
        \Proj{1}
      )
    : \ICont{2}
  \).
\end{problem}
  \begin{construction}
    Our goal is to provide some
    \(
      \alpha : \Cart{ \Subst{F'}{\Der{\mu F}} }{ \Der{\mu F} }
    \).
    First note that \( F' \) is a container in two parameters,
    and substitution into the second replaces \( \Proj{1} \), i.e.\@ there is an equivalence
    \[
      \gamma_{\Highlight{Y}} :
      F'[\Highlight{Y}]
        \mathrel{\CartEquiv}
      { \Der_0{F}[ {\mu F} ] } \CPlus ( {\Der_1{F}[ {\mu F} ]} \CTimes \Highlight{Y} )
    \]
    for any choice of \( \Highlight{Y} : \ICont{1} \).
    In particular, for \( \Highlight{Y} \JudgeEq \Der(\mu F) \),
    the codomain of \( \gamma_{\Der{\mu F}} \) is exactly the domain of the parametrized chain rule (cf.~\autoref{binary-chain-rule}).
    Hence, we define \( \alpha \) as the following composite:
    \begin{equation*}
      \begin{tikzcd}[column sep=large]
        {F'[\Highlight{\Der(\mu F)}]}
          &
        {\Der(\mu F)}
          \\
        {
          { \Der_0{F}[ {\mu F} ] }
            \CPlus
          (
            {\Der_1{F}[ {\mu F} ]}
              \CTimes
            \Highlight{\Der(\mu F)}
          )
        }
          &
        \Der(F[\mu F])
        \ar[from=1-1, to=2-1, multimap-multimap, "\gamma_{\Highlight{\Der(\mu F)}}"{swap}]
        \ar[from=2-1, to=2-2, -multimap, "\Chain_{F,{\mu\!F}}"{swap}]
        \ar[from=2-2, to=1-2, multimap-multimap, "\Der(\In_F)"{swap}]
        \ar[from=1-1, to=1-2, dashed, -multimap, "\alpha"]
      \end{tikzcd}
    \end{equation*}
    Lastly, let \( \MuRule_F \DefEq \Rec_{F'}(\alpha) : \Cart{ \mu F' }{ \Der{(\mu F)} } \).
  \end{construction}

One advantage of defining the \( \mu \)-rule by recursion is that it highlights the dependency on the chain rule.
In fact, we will spend the remainder of this chapter proving that \( \MuRule_F \) is strong
if and only if \( \Chain_{F,\mu F} \) is~(\autoref{strong-mu-rule-iff-strong-chain-rule}).
First, observe that recursion reflects equivalence, in the following sense:
\begin{lemma}\label{is-equiv-from-mu-rec}
  Let \( \alpha : \Cart{\Subst{F}{G}}{G} \).
  If \( \Rec_{F}(\alpha) \) is an equivalence, then so is \( \alpha \).
  \begin{proof}
    Substitution \( \Subst{F}{-} \) preserves equivalences,
    so by 3-for-2 for equivalences of containers, \( \alpha \) on the right is an equivalence:
    \[
      \begin{tikzcd}[column sep=huge]
        \Subst{F}{\mu F} & \Subst{F}{G} \\
        {\mu F}          & G
        \ar[from=1-1, to=1-2, multimap-multimap, "\Subst{F}{\Rec_F(\alpha)}"]
        \ar[from=1-1, to=2-1, multimap-multimap, "\In_F"{swap}]
        \ar[from=2-1, to=2-2, multimap-multimap, "\Rec_F(\alpha)"{swap}]
        \ar[from=1-2, to=2-2, -multimap, "\alpha"]
      \end{tikzcd}
      \qedhere
    \]
  \end{proof}
\end{lemma}

Thus, a strong \( \mu \)-rule for some container necessarily implies a strong chain rule between \( F \) and \( \mu F \):
\begin{proposition}\label{strong-chain-rule-from-strong-mu-rule}
  Let \( F : \ICont{I + 1} \).
  If \( \MuRule_F \) is an equivalence, then so is \( \Chain_{F,{\mu F}} \).
  \begin{proof}
    Assume that \( \MuRule_F \) is an equivalence.
    In \autoref{mu-rule}, \( \MuRule_F \) is defined by recursion from some \( \alpha \),
    hence \( \alpha \) is an equivalence by \autoref{is-equiv-from-mu-rec}.
    But \( \alpha \) is just \( \Chain_{F, {\mu F}} \) wedged between equivalences,
    so the latter is an equivalence.
  \end{proof}
\end{proposition}

We would like to establish the converse to this as well, but cannot do so directly,
as the converse of \autoref{is-equiv-from-mu-rec} does not hold:
in general, there are fixpoints \( \varphi : \Subst{F}{G} \CartEquiv G \) such that \( \mu F \) and \( G \) are not equivalent as containers.
If ordered with respect to embeddings, however, \( \mu F \) is the least pre-fixpoint of \( \Subst{F}{\Blank} \):
\begin{lemma}\label{is-embedding-mu-rec}
  If \( \alpha : \Cart{\Subst{F}{G}}{G} \) is an embedding of containers,
  then so is \( \Rec_F(\alpha) : \Cart{\mu F}{G} \).
  \begin{proof}
    This is a direct application of the next result (\autoref{is-embedding-W-rec}) to the shape map \( \Rec_{F^{\prime}}(\alpha)_{\Sh} : \W_{S}(P) \to T \).
  \end{proof}
\end{lemma}

Fix some \( S : \Type \) and a family \( P : S \to \Type \).
Recall that non-dependent recursion on \( \W_{S}(P) \) into a type \( A \) is given by
a map
\begin{align*}
  \WRec &: \bigl(\bigl( { \textstyle \sum_{s : S} (P(s) \to A) } \bigr) \to A \bigr) \to (\W_{S}(P) \to A)
\end{align*}
such that for any \( h : \left(\sum_{s : S} (P(s) \to A)\right) \to A \),
\begin{align*}
  \WRec&(h, \Sup(s, f)) \JudgeEq h(s, \WRec(h) \circ f).
\end{align*}
We can characterize the fibers of \( \WRec(h) \) when \( h \) is an embedding:
\begin{problem}\label{prob:wrec-fiber-equiv}
  Let \( A : \Type \) and \( { h : ( \sum_{s : S} (P(s) \to A) ) \hookrightarrow A } \).
  For all \( s : S \) and \( f : P(s) \to \W_{S}(P) \),
  define an equivalence
  \[
    \Fiber_{\WRec(h)}(\WRec(h, \Sup(s, f))) \simeq \prod_{p : P(s)} \Fiber_{\WRec(h)}(\WRec(h, f(p)))
  \]
  characterizing the fibers over the image of \( \WRec(h) : \W_{S}(P) \to A \).
\end{problem}
\begin{construction}
  Abbreviate \( h^* \DefEq \WRec(h) \) and let \( s : S \), \( f : P(s) \to \W_{S}(P) \).
  By definition, we have
  \begin{align*}
    \Fiber_{h^*}(h^*(\Sup(s, f)))
      &\simeq
      \smashoperator{\sum_{w : \W_{S}(P)}}\, h^*(w) = h(s, h^* \circ f)
    \intertext{
      Unfolding \( w \) via \( \WIn : \sum_{s : S } (P(s) \to \W_{S}(P)) \simeq \W_{S}(P) \), we obtain
    }
      &\simeq
      \adjustlimits\sum_{s' : S} \sum_{\smash{f'} : P(s') \to \W_{S}(P)}
        h^*(\Sup(s', f')) = h(s, h^* \circ f)
    \intertext{
      By the computation rule for \( \W \), this is equivalent to
    }
      &\simeq
      \adjustlimits\sum_{s' : S} \sum_{\smash{f'} : P(s') \to \W_{S}(P)}
        h(s', h^* \circ f') = h(s, h^* \circ f)
    \intertext{
      Since \( h \) is an embedding, it can be cancelled from the path type
    }
      &\simeq
      \adjustlimits\sum_{s' : S} \sum_{\smash{f'} : P(s') \to \W_{S}(P)}
        (s', h^* \circ f') = (s, h^* \circ f)
    \intertext{
      Breaking up the path of pairs into a pair of paths, we rearrange:
    }
      &\simeq
      \adjustlimits\sum_{(s\mathrlap{'} , q) : \Singl(s)} \sum_{\smash{f'} : P(s') \to \W_{S}(P)}
        h^* \circ f' =_{q} h^* \circ f
    \intertext{
      Contracting \( \Singl(s) \) and applying function extensionality to the right,
      we are left with
    }
      &\simeq
      \sum_{\smash{f'} : P(s) \to \W_{S}(P)}
        \prod_{p : P(s)}
        h^*(f'_p) = h^*(f_p)
    \intertext{
      Distributing \( \Sigma \) and \( \Pi \),
      we get
    }
      &\simeq
        \prod_{p : P(s)}
          \sum_{w : \W_{S}(P)}
          h^*(w) = h^*(f_p) \\
      &\simeq
        \prod_{p : P(s)}
          \Fiber_{h^*}(h^*(f_p))
  \end{align*}
  which finishes the construction of the desired equivalence.
\end{construction}

Consequently, \( \WRec \) preserves embeddings.
\begin{lemma}\label{is-embedding-W-rec}
  For all \( A : \Type \),
  if the function \( { h : ( \sum_{s : S} (P(s) \to A) ) \to A } \) is an embedding,
  then so is \( \WRec(h) : \W_{S}(P) \to A \).
  \begin{proof}
    Again let \( h^* \DefEq \WRec(h) \).
    It suffices to show that fibers over the image of \( h^* \) are propositions,
    i.e.\@ \( \prod_{w : \W_{S}(P)} \Op{isProp}(\Fiber_{h^*}(h^*(w)))  \).
    We do so by \( \W \)-induction.
    Let \( s : S \), \( f : P(s) \to \W_{S}(P) \).
    By induction, we can assume
    \[
      \eta : \prod_{p : P(s)} \Op{isProp}(\Fiber_{h^*}(h^*(f_p)))
    \]
    By \autoref{prob:wrec-fiber-equiv}, there is an equivalence over \( h^*(\Sup(s, f)) \), namely
    \begin{align*}
      \Fiber_{h^*}(h^*(\Sup(s, f)))
        &\simeq
          \prod_{p : P(s)}
            \Fiber_{h^*}(h^*(f_p))
    \end{align*}
    By the induction hypothesis \( \eta \), the type on the right is a proposition.
    Hence, all fibers over the image of \( h^* \) are propositions,
    and \( h^* \) is an embedding.
  \end{proof}
\end{lemma}

This tells us that \( \MuRule_F \) embeds \( \mu F' \) as a subcontainer inside of \( \Der(\mu F) \):
\begin{lemma}\label{is-embedding-mu-rule}
  For all \( F : \ICont{I + 1} \), \( \MuRule_F : \Cart{\mu F'}{\Der(\mu F)} \) is an embedding.
  \begin{proof}
    By the parametrized analogue of \autoref{is-embedding-chain-rule}, the chain rule is an embedding.
    This lifts to \( \alpha \), and from there to \( \MuRule_F \JudgeEq \Rec_F(\alpha) \) by \autoref{is-embedding-mu-rec}.
  \end{proof}
\end{lemma}

From this, we can conclude that a strong chain rule alone is enough to prove a strong \( \mu \)-rule:
\begin{proposition}\label{strong-chain-rule-to-strong-mu-rule}
  Let \( F : \ICont{I + 1} \).
  If  \( \Chain_{F,{\mu F}} \) is an equivalence, then so is \( \MuRule_F \).
  \begin{proof}
    Since \( \MuRule_F \) is always an embedding (\autoref{is-embedding-mu-rule}),
    it suffices to show that its shape map is a surjection.
    In fact, it is a split surjection: it has a section, hence all of its fibers are inhabited, not just merely inhabited.
    Assume \( \Chain_{F,{\mu F}} \) to be an equivalence.
    By \autoref{strong-binary-chain-rule-iff-is-equiv-sigma-isolate}, this is equivalent to isolated pairs
    \( (p_1, \bar{w}) \)
    having isolated components \( p_1 : P_1(s) \) and \( \bar{w} : \WPath(f{p_1}) \),
    for all \( s : S \) and \( f : P_1(s) \to \W_{\!S}(P_1) \).
    This property is exactly what is needed to define a section to the shape map.
  \end{proof}
\end{proposition}

Thus, whether the \( \mu \)-rule is strong depends only on the chain rule:
\begin{theorem}\label{strong-mu-rule-iff-strong-chain-rule}
  For any container \( F : \ICont{I + 1} \), the following are equivalent:
  \begin{enumerate}
    \item \( \MuRule_F \) is an equivalence.
    \item \( \Chain_{F,{\mu F}} \) is an equivalence.
  \end{enumerate}
  \begin{proof}
    One direction is \autoref{strong-chain-rule-from-strong-mu-rule},
    the other is \autoref{strong-chain-rule-to-strong-mu-rule}.
  \end{proof}
\end{theorem}

With this at hand, we can give a proof that the \( \mu \)-rule is strong for discrete containers
that factors solely through the properties of the chain rule.
First, note that \( \WPath \) preserves discreteness:
\begin{lemma}
  If \( B, C : A \to \Type \) are families of discrete types,
  then \( \WPath_{A,B,C} \) is a discrete family.
  \begin{proof}
    By induction on \( \W_{A}(B) \) and unfolding via \( \WPathIn \).
  \end{proof}
\end{lemma}
This implies that discrete containers are closed under \( \mu \):
\begin{lemma}\label{mu-discrete}
  If \( F : \ICont{I + 1} \) is a discrete container, then so is \( \mu F \).
\end{lemma}
Hence, we obtain a strong \( \mu \)-rule for discrete containers:
\begin{proposition}[note={\cite[{Proposition 8.1}]{AbbottEtAl2005DataDifferentiating}}]
  If \( F : \ICont{I + 1} \) is discrete, then \( \MuRule_F \) is an equivalence.
  \begin{proof}
    By \autoref{strong-mu-rule-iff-strong-chain-rule}, it suffices to show that \( \Chain_{F,\mu F} \) is an equivalence.
    But \( \mu F \) is discrete (\autoref{mu-discrete}), and the chain rule between discrete containers is strong (\autoref{discrete-strong-binary-chain-rule}).
  \end{proof}
\end{proposition}

\section{Conclusion}

We have seen that extending a derivative operation to \( \Type \)-valued containers is possible:
If the obstruction to a general derivative is that positions are not necessarily discrete,
simply restrict a derivative to the subtype of those positions that are!
This generalization does not come for free, however.
Although it is universal and retains basic properties, the chain rule is no longer in general invertible,
and neither is the \( \mu \)-rule.
In fact, a global chain rule is impossible, and at least constructively taboo when restricted to sets.

Interestingly, the status of the \( \mu \)-rule is somewhat weaker:
In the case of the chain rule the contradiction arises because we are free to apply it to an arbitrary pair of containers;
in \autoref{globally-discrete-iff-strong-chain-rule} the strategy is to pick an \enquote{adversarial}
container \( G \) which forces the positions of an arbitrary container \( F \) to be discrete.
In \autoref{strong-mu-rule-iff-strong-chain-rule}, however, the constraints are tighter
---
there is one degree of freedom (the container \( F \)),
and whether the rule is strong depends on that and \( \mu F \),
which is canonically derived from \( F \).
We would like to know:
Is a strong \( \mu \)-rule a constructive taboo in the same way that a strong chain rule is?
We currently do not know of any container \( F \) for which \( \MuRule_F \) fails to be strong.
At the same time, proving that it is always strong may require more advanced proof techniques,
such as the generalized \( \WPath \)-induction of~\cite[Lemma~12]{DamatoEtAl2025FormalisingInductiveCoinductive}.

In the future we would also like to work out the behavior of \( \Der \) on fixpoints other than \( \mu \).
In their original work~\cite{AbbottEtAl2005DataDifferentiating}, \citeauthor{AbbottEtAl2005DataDifferentiating} derive a fixpoint rule also for the greatest fixpoint, \( \nu F \).
Recently, \citeauthor{DamatoEtAl2025FormalisingInductiveCoinductive} show that \( \Type \)-valued container functors preserve both least and greatest fixpoints~\cite{DamatoEtAl2025FormalisingInductiveCoinductive}.
In either work, many of the intermediate lemmata hold for arbitrary fixpoints,
and the same is true for a number of our results:
The construction of the lax chain rule in \autoref{lax-chain-rule}, for example, only depends on the property of \( \In_F : \Cart{\Subst{F}{\mu F}}{\mu F} \) being \emph{some}
\( \Subst{F}{\Blank} \)-fixpoint, not necessarily the least fixpoint.
We could thus give for any \( \varphi : \CartIso{\Subst{F}{\varphi F}}{\varphi F} \) a \enquote{\( \varphi \)-rule}
as an embedding \( \operatorname{\Op{\varphi-rule}} : \Cart{\mu F'}{\Der(\varphi F)} \) in which
\(
  F' \DefEq
    { \Wk{\Der_0{F}[ {\varphi F} ]} }
      \CPlus
    (
      \Wk{\Der_1{F}[ {\varphi F} ]}
        \CTimes
      \Proj{1}
    )
\).

In another direction, we are curious whether we can describe more than just one-hole contexts.
\Citeauthor{AbbottEtAl2003DerivativesContainers} define \emph{linear exponentials} \cite[{Def.~4.1}]{AbbottEtAl2003DerivativesContainers}:
for containers \( F \) and \( H \), the linear exponential of \( [H, F] \) exists if there is a universal morphism from \( \Blank \CTimes H \) to \( F \).
They show that \( \Der{F} = [\Id, F] \) if \( F \) is discrete.
The adjunction of \autoref{derivative-adjunction-sets} is such a universal morphism, natural in all \( F \),
hence \( \Der = [\Id, \Blank] \) as functors.
Iterating the adjunction, we can describe \( n \)-hole contexts as linear exponentials \( \Der^n = [\CYo\Fin{n}, \Blank] \).
We conjecture that \( [\CYo(H), (\MkCont{S}{P})] \) exists for arbitrary \( H : \Type \):
for any embedding \( f : H \hookrightarrow P_s \) whose fibers are decidable,
the complement \( P_s -_f H \DefEq \sum_{p : P_s} \neg \Fiber_f(p) \) satisfies, among other things, \( (P_s -_f H) + H \simeq P_s \).
The construction of this equivalence does not depend on discreteness or the truncation level of \( H \).
This indicates that a type of contexts with \enquote{\( H \)-shaped} holes may exist even if \( H \) is a higher type.

\paragraph{Acknowledgments.}
This work was supported by the Estonian Research Council grant PSG749.

\printbibliography

\end{document}